\documentclass[iop]{emulateapj}

\usepackage{natbib}
\usepackage{amsmath, amssymb}
\usepackage{times}
\usepackage{apjfonts, graphicx, graphics, mathtools}
\usepackage[breaklinks,colorlinks,urlcolor=blue,citecolor=blue]{hyperref}
\usepackage[all]{hypcap}
\usepackage{aas_macros}
\usepackage{subfigure}
\usepackage{multirow}
\usepackage{hhline}

\newcommand{\tctff}{{t$_{\rm c}$/t$_{\rm ff}$}}
\newcommand{\mm}{{$M_{\rm mol}$}}

\shorttitle{Molecular clouds in early-type galaxies}
\shortauthors{Iu. V. Babyk et al.}

\begin{document}

\title{Origins of molecular clouds in early-type galaxies}
\author{Iu. V. Babyk$^{1,2,3,\ast}$}
\author{B.~R. McNamara$^{1,2,4}$}
\author{P. D. Tamhane$^{1,2}$}
\author{P.~E.~J. Nulsen$^{5,6}$}
\author{H.~R. Russell$^{7}$}
\author{A.~C. Edge$^{8}$}

\affil{
    $^{1}$ Department of Physics and Astronomy, University of Waterloo, Waterloo, ON, N2L 3G1, Canada \\
    $^{2}$ Waterloo Institute for Astrophysics, University of Waterloo, Waterloo, ON, N2L 3G1, Canada \\
    $^{3}$Department for Extragalactic Astronomy and Astroinformatics, Main Astronomical Observatory of the National Academy of Sciences of Ukraine, \\ 27 Zabolotnoho str., 03143, Kyiv, Ukraine \\
    $^{4}$ Perimeter Institute for Theoretical Physics, 31 Caroline str. N, Waterloo, ON, N2L 2Y5, Canada \\
    $^{5}$ Harvard-Smithsonian Center for Astrophysics, 60 Garden Street, Cambridge, MA 02138, USA \\
    $^{6}$ ICRAR, University of Western Australia, 35 Stirling Hwy, Crawley, WA 6009, Australia \\
    $^{7}$ Institute of Astronomy, Madingley Road, Cambridge CB3 0HA, UK \\
    $^{8}$ Department of Physics, University of Durham, South Road, Durham DH1 3LE, United Kingdom \\
    \\
}

\begin{abstract}
\hspace{0.5cm} We analyze $Chandra$ observations of the hot atmospheres of 40 early spiral and elliptical galaxies. Using new temperature, density, cooling time, and mass profiles, we explore relationships between their hot atmospheres and cold molecular gas.  Molecular gas mass correlates with atmospheric gas mass and density over four decades from central galaxies in clusters to normal giant ellipticals and early spirals. The mass and density relations follow power laws: $M_{\rm mol}  \propto M_{\rm X}^{1.4\pm0.1}$ and $M_{\rm mol}  \propto n_{\rm e}^{1.8\pm0.3}$, respectively, at 10 kpc. The ratio of molecular gas to atmospheric gas within a 10 kpc radius lies between $3\%$ and $10\%$ for early-type galaxies and between $3\%$ and $50\%$ for central galaxies in clusters.
Early-type galaxies have detectable levels of molecular gas when their atmospheric cooling times falls below $\sim \rm Gyr$ at a radius of 10 kpc.
A similar trend is found in central cluster galaxies. We find no relationship between the ratio of the cooling time to free fall time, $t_{\rm c}/t_{\rm ff}$, and the presence or absence of molecular clouds in early-type galaxies. The data are consistent with much of the molecular gas in early-type galaxies having condensed from their hot atmospheres. 
\end{abstract}


\altaffiltext{*}{
    \href{mailto:ibabyk@uwaterloo.ca, babikyura@gmail.com}{ibabyk@uwaterloo.ca, babikyura@gmail.com}
}

\section{Introduction}\label{sec:intro}

Elliptical galaxies were historically thought to be devoid of gas.  However, modern studies have shown that many early-type galaxies (ETGs) contain gas at broad range of temperatures.  Most abundant are the hot, ten million degree atmospheres observed in X-rays \citep{Forman:85, Trinchieri:85, Mathews:03, Kim:15}. Their X-ray luminosity scales with temperature as $L_{\rm X} \propto T^{4.5}$ \citep{Boroson:10, Kim:13, Babyk:17scal}, indicating a strong link between atmospheric luminosity and halo mass \citep{Forbes:17}. Their hot atmospheres likely formed primarily from cool gas that accreted early and was heated to the virial temperature by shocks, and by gas expelled from stars \citep{Goulding:16}.

Ellipticals and early spirals contain modest levels of molecular and atomic hydrogen.  Dust is commonly  seen in the form of clouds and lanes near to the nucleus \citep{Sadler:85, Goudfrooij:94, Vandokkum:95}.  Dust is usually associated with larger quantities of cold atomic and molecular gas \citep{Combes:07}. Thirty to forty percent of early-type galaxies contain detectable levels of molecular gas \citep{Combes:07, Salome:11, Young:11}. Infrared and radio observations indicate that a minority of systems with relatively large amounts of molecular gas form stars at rates of $\lesssim 0.1~\rm M_\odot ~yr^{-1}$ or so \citep{Combes:07, Shapiro:10, Ford:13}.   

Neutral hydrogen is less abundant than molecular hydrogen in early-type galaxies. In a study of 33 ETGs from the SAURON sample, HI is detected at a level of $10^6~\rm M_\odot$ in 2/3 of field galaxies.  Detections plunge to 10\% in cluster galaxies \citep{Oosterloo:10, Pulatova:15}. Like galaxies detected in CO, fast rotating galaxies are more likely to harbour HI than slow rotators. Detection is a strong function of environment. Isolated galaxies are more likely to harbour HI than those in clusters. This implies that galaxies are either stripped of their HI,  or they are unable to accrete material from their surroundings within clusters. Or both.  However, the prevalence of dynamically young HI structures connecting to disks suggests that at least some gas is accreted externally \citep{Oosterloo:10}.

The origin of cold gas in early-type galaxies has been debated for decades.  Its origin is key to understanding how early-type galaxies formed and co-evolved with massive nuclear black holes, why so few are experiencing significant levels of star formation, and if or how they are maintained by AGN feedback \citep{Kormendy:13}. The absence of correlation between molecular gas mass and the host's stellar mass is inconsistent with mass loss from stars in the galaxy \citep{Combes:07}.  However, the similarity between nebular gas metallicities, which are likely associated with molecular clouds, and stellar metallicity in four early-type galaxies is consistent with an internal origin including stellar mass loss and/or cooling from the hot atmosphere, but is inconsistent with having accreted from other galaxies \citep{Griffith:18}. On the other hand, counter rotation between stars and cold gas found in some systems and the prevalence of molecular gas in fast rotators \citep{Young:11} may imply an external origin, perhaps due to mergers of gas-rich companions.   

Taking a fresh approach, \citet{Werner:14} studied [CII]$\lambda 157 \mu \rm m$,  [OI]$\lambda 63 \mu \rm m$, and [OIb]$\lambda 145 \mu \rm m$ emissions in eight giant ellipticals and BCGs.  They compared this emission to their atmospheric properties. This microwave line radiation emerges from warm molecular, neutral or ionized gas at a temperature of $\sim 100$ K that is associated with nebular emission. Comparing the molecular gas emission to their atmospheric properties, Werner found stronger emission lines in systems with lower atmospheric gas entropy and shorter cooling times. In newer work, \citet{Werner:18new} found a similar connection between nebular emission and cooling hot atmospheres,  indicating a close connection between the hot atmosphere and molecular gas.

In this respect, early-type galaxies appear similar to brightest cluster galaxies, but scaled down to much lower molecular gas masses.  Brightest cluster galaxies (BCGs), the elliptical-like galaxies at the centers of clusters, often contain reservoirs of cold molecular gas with masses lying between $10^9~\rm M_\odot$ to several $10^{11}~\rm M_\odot$  \citep{Edge:01, Salome:03}. The molecular gas masses correlate with nebular emission \citep{Edge:01}, which is known to be  associated with so-called cool cores, the cooling central regions of galaxy cluster atmospheres \citep{Heckman:89, Hu:85, Crawford:99, Cavagnolo:08, McDonald:10}. Like nebular emission  and star formation \citep{Rafferty:08}, molecular gas is known to form in systems when the cooling time of the hot atmosphere falls below a cooling time threshold of $\sim 10^9$ yr within an altitude of approximately 10 kpc \citep{Pulido:17}. The existence of this threshold, in addition to other trends between molecular gas mass and atmospheric mass and density \citep{Pulido:17}, indicate that molecular gas condensed out the hot atmospheres. 

Atacama Large Millimetre Array (ALMA) has observed nearly a dozen central galaxies in groups and clusters \citep{David:14, McNamara:14, Russell:14, Russell:16, Russell:17a, Russell:17b, Tremblay:16, Vantyghem:16, Vantyghem:17, Simionescu:18}.  ALMA images show that much of the molecular gas lies outside of the nucleus in most systems. The gas appears in wisps and filaments traveling at slower velocities than the average speeds of the stars. Circumnuclear disks are rare. In many systems the molecular gas lies below and appears to have been drawn-up behind the rising X-ray bubbles, as was noted in Perseus a decade ago \citep{Fabian:03, Hatch:06, Salome:11, Hlavacek:15, Yang:16}.  This suggests two scenarios: preexisting molecular gas is drawn up by the bubbles, or low entropy, keV atmospheric gas is drawn up behind the bubbles and cools in their wakes \citep{Salome:11}. 

Recognizing the difficulty lifting high density molecular clouds, \citet{McNamara:16} proposed that the keV gas become thermally unstable and cools behind the bubbles when the ratio of the cooling time to the timescale to fall back to the galaxy approaches unity, $t_{\rm c}/t_{\rm I} \lesssim 1$. The infall time is determined by the lesser of the free fall and terminal speeds of the thermally unstable gas. This model emerged from ALMA observations and early observations of the Perseus cluster \citep{Russell:17a}. Furthermore, $Chandra$ observations have shown that atmospheric gas is lifted  behind rising X-ray bubbles with displaced masses comparable to the molecular gas masses of their hosts \citep{Fabian:02, Simionescu:10, Gitti:11, Werner:12, Kirkpatrick:15}. Thermally unstable cooling in uplifted gas is seen in  simulations \citep{Revaz:08, Li:14}. 

This model implies a tight link between molecular gas and AGN feedback where the same process that heats and stabilizes the atmosphere creates its own cold fuel that feeds the AGN \citep{Gaspari:12}.  The self-generation of cold gas would sustain radio mode feedback \citep{McNamara:16}.  

How molecules residing in the atmospheres of early-type galaxies formed is poorly understood.  They may have arrived externally via mergers or inflowing cold filaments. They may have condensed internally from a combination of thermalized stellar ejecta and their hot atmospheres. Molecular cloud formation in disk galaxies is thought to be catalyzed primarily on dust grains \citep{LeBourlot:12}.  However, unshielded dust may be sputtered away in the harsh atmospheric environment of early-type galaxies \citep{Draine:79}, making it difficult to survive a merger and perhaps more difficult to form from cooling atmospheres.  Nevertheless, despite the harsh atmospheric conditions, dust features are prevalent in early-type galaxies, particularly in systems rich in molecular clouds \citep{Russell:19, Werner:19}. Upwards of $10^{10} ~\rm M_\odot$ of molecular gas is observed in BCGs \citep{Edge:01} which cannot have originated in mergers \citep{Pulido:17}. Molecular clouds are apparently able to form and survive in these systems despite harsh atmospheric environments.

Here, we analyze the  thermodynamic properties of the hot atmospheres of 40 early-type galaxies from $Chandra$ archival data. Temperature, density, entropy, cooling time, mass, and free-fall time profiles are measured. The onset of thermally unstable cooling is investigated in the context of the molecular gas reservoirs and their role in the AGN feedback.  Archival ALMA data are used to examine the molecular gas structure in some systems. We combine this study with the \citet{Pulido:17} sample of cluster BCGs with molecular gas to study trends between normal ellipticals and central galaxies in clusters over a large range in halo mass and molecular gas mass.

The paper is organized as follows. In Section~\ref{sec:2} we describe our sample and data reduction of the $Chandra$ observations. In Section~\ref{sec:3} we present physical characteristics of the targets including temperature, density, entropy, cooling, mass, and free-fall time profiles. In Section~\ref{sec:4} we study \tctff\ ratio to explore recently presented theoretical models of thermally unstable gas. In Section~\ref{sec_molec} we present correlations between molecular gas mass and X-ray gas properties. Conclusions are presented in Section~\ref{sec_conc}.

We adopt the following cosmological parameters: $H_0$ = 70 km s$^{-1}$ Mpc$^{-1}$, $\Omega_{\Lambda}$ = 0.7, $\Omega_{M}$ = 0.3. 

\begin{table*}
\caption{Early-type galaxies/faint groups sample.}\label{tab1}
\centering
\begin{tabular}{lcccccccccccc}
\hline
 && \\
Name      & R.A. & Decl. & ObsIDs & Exposure & Type & BCG  &  $z$     &  $D_{\rm A}$ &  $D_{\rm L}$  &  $N_{\rm H}$ & 1.4 GHz Flux \\
          &  (J2000)   & (J2000)    &       &    ks     &      &         &          &   Mpc  &  Mpc    &  10$^{20}$ cm$^{2}$ & mJy \\
          & (2) & (3) & (4) & (5) & (6) & (7) &(8) & (9) & (10) & (11) & (12) \\
&& \\
\hline
&&\\
IC1262  & 17:33:02.022 & +43:45:34.51 & 6949, 7321, 7322 & 36.02, 34.98, 35.17 & E & $\surd$ & 0.032649 & 133.0 & 141.8 & 2.47 & 16.4$\pm$1.0 \\
IC1459  & 22:57:10.608 & -36:27:43.997 & 2196 & 45.14 & E3 & & 0.006011 & 25.503 & 25.8 & 1.19 & 1279.7$\pm$45.2 \\
IC4296  & 13:36:39.053 & -33:57:57.30 & 2021, 3394 & 19.27, 20.78 & E &  & 0.012465 & 52.358 & 53.7 & 4.11 & 546.6$\pm$17.8 \\
NGC315  & 00:57:48.883 & +30:21:08.812 & 4156 & 39.49 & E &         & 0.016485 & 68.816 & 71.1 & 5.87 & 772.1$\pm$25.3 \\
NGC499  & 01:23:11.459 & +33:27:36.30 & 10536 & 18.33 & E & & 0.014673 & 61.423 & 63.2 & 5.26 & 62.7$\pm$1.9 \\
        & && 10865 & 5.12 & \\
        & && 10866 & 8.01 & \\
        & && 10867 & 7.02 & \\
NGC507  & 01:23:39.950 & +33:15:22.22 & 317 & 40.30 & E & $\surd$ & 0.016458 & 68.706 & 71.0 & 5.32 & 61.7$\pm$2.5 \\
NGC533  & 01:25:31.432 & +01:45:33.57 & 2880 & 28.40 & E3 & & 0.018509 & 77.025 & 79.9 & 3.12 & 28.6$\pm$1.0 \\
NGC708  & 01:52:46.482 & +36:09:06.53 &  2215, 7921 & 28.75, 108.63   & E    & $\surd$ & 0.016195 & 67.635 & 69.8  &  5.37 & 65.7$\pm$2.3\\
NGC720  & 01:53:00.523 & -13:44:19.25 & 7372 & 49.13 & E5 &    & 0.005821 & 24.704 & 25.0 & 1.55 & 96.2$\pm$3.4\\
        & && 7062 & 22.12 & \\
        & && 8448 & 8.06 & \\
        & && 8449 & 18.91 & \\
NGC741  & 01:56:20.959 & +05:37:43.77 & 2223 & 28.14 & E0 &         & 0.018549 & 77.186 & 80.1 & 4.47 & 478.8$\pm$16.2 \\
NGC1316 & 03:22:41.789 & -37:12:29.52 & 2022  & 21.21  & E &         & 0.005871 & 24.914 & 25.2  & 1.92 & 254.7$\pm$9.9\\
NGC1332 & 03:26:17.321 & -21:20:07.33 & 2915, 4372  & 4.10, 16.38  & S0  &         & 0.005084 & 21.601 & 21.8  & 2.29 & 4.6$\pm$0.5 \\
NGC1399 & 03:38:29.083 & -35:27:02.67 & 9530  & 56.98    & E1   & $\surd$ & 0.004753 & 20.205 & 20.4  &  1.31 & 208.0$\pm$6.9\\
NGC1404 & 03:38:51.917 & -35:35:39.81 & 16233 & 91.94  & E1  & & 0.006494 & 27.531 & 27.9  & 1.35 & 3.9$\pm$0.6 \\
        & && 16231 & 56.09 &\\
        & && 16232 & 64.03 &\\
        & && 16234 & 84.64 &\\
NGC1407 & 03:40:11.904 & -18:34:49.36 & 14033 & 50.26    & E0   &         & 0.005934 & 25.179 & 25.5  &  5.41 & 87.7$\pm$3.5 \\
NGC1550 & 04:19:37.921 & +02:24:35.58 & 5800, 5801 & 44.55, 44.45 & E2   & $\surd$ & 0.012389 & 52.045 & 53.3 & 11.2  & 16.6$\pm$1.6 \\
NGC3091 & 10:00:14.125 & -19:38:11.32 & 3215 & 27.34 & E3 & $\surd$ & 0.013222 & 55.473 & 56.9 & 4.75 & 2.5$\pm$0.5 \\
NGC3923 & 11:51:01.783 & -28:48:22.36  & 9507  & 80.90  & E4  &         & 0.005801 & 24.620 & 24.9 & 6.29 & 31.2$\pm$1.1 \\
NGC4073 & 12:04:27.059 & +01:53:45.65 & 3234 & 25.76 & E  & $\surd$ & 0.019584 & 81.364 & 84.6 & 1.90 & 17.1$\pm$1.0 \\
NGC4104 & 12:06:38.910 & +28:10:27.17 &  6939 & 34.86    & S0  & $\surd$ & 0.028196 & 115.60  & 122.2 &  1.68 & 7.3$\pm$0.5 \\
NGC4125 & 12:08:06.017 & +65:10:26.88  & 2071   & 52.97 & E6  &         & 0.004523 & 19.234 & 19.4 & 1.86 & 24.9$\pm$1.2 \\
NGC4261 & 12:19:23.216 & +05:49:29.695 &  9569 & 102.24   & E2 &         & 0.007378 & 31.236 & 31.7  &  1.56 & 4066.7$\pm$124.0 \\
NGC4325 & 12:23:06.672 & +10:37:17.05 & 3232 & 28.30 & E4 & $\surd$ & 0.025714 & 105.80 & 111.3 & 2.18 & 4.1$\pm$0.5 \\
NGC4374 & 12:25:03.743 & +12:53:13.19  & 5908, 6131 & 44.04, 35.81  & E1 &           & 0.003392 & 16.422 & 18.1 & 2.58 & 27.5$\pm$2.0 \\
NGC4382 & 12:25:24.053 & +18:11:27.89  & 2016 & 29.33  & S0 &          & 0.002432 & 16.265 & 17.7 & 2.51 & 8.0$\pm$0.5  \\
NGC4472 & 12:29:46.798 & +08:00:01.48 & 11274 & 39.67    & E2   &         & 0.003272 & 15.621 & 17.1  &  1.65 & 219.9$\pm$7.8\\
NGC4552 & 12:35:39.8 & +12:33:23  & 13985 & 49.41 & E  &           & 0.001134 & 15.523 & 16.1 & 2.56 & 100.1$\pm$3.0 \\
        & && 14358 & 49.41 & \\
        & && 14359 & 47.11 & \\
NGC4636 & 12:42:49.867 & +02:41:16.01 & 3926, 4415  & 67.26, 66.17    & E0   &         & 0.003129 & 13.335 & 13.4  &  1.83 & 77.8$\pm$2.8 \\
NGC4649 & 12:43:40.008 & +11:33:09.40 & 8182, 8507  & 45.87, 15.73    & E2   &         & 0.003703 & 15.767 & 15.9  &  2.13 & 29.1$\pm$1.3 \\
NGC4696 & 12:48:49.277 & -41:18:39.92 & 1560  & 21.20 & E1 & $\surd$ & 0.009867 & 41.613 & 42.4 & 8.07 & 24.4$\pm$6.0 \\
NGC4782 & 12:54:35.698 & -12:34:06.92 & 3220 & 49.33 & E0 &          & 0.015437 & 64.545 & 66.6 & 3.56 & 3.1$\pm$0.4 \\
NGC5044 & 13:15:23.969 & -16:23:08.00 & 17195 & 77.01    & E0   & $\surd$ & 0.00928  & 39.173 & 39.9  &  5.03 & 34.7$\pm$1.1  \\
        &          &          & 17196 & 85.80    &      &         \\
        &          &          & 17653 & 32.46    &      &         \\
        &          &          & 17654 & 24.01    & \\
        &          &          & 17666 & 82.79    & \\
NGC5353 & 13:53:26.7 & +40:16:59.0 & 14903 & 37.20 & S0 &            & 0.007755 & 32.813 & 33.3 & 0.98 & 40.5$\pm$1.3 \\
NGC5813 & 15:01:11.265 & +01:42:07.09 & 12952 & 140.00 & E1 &        & 0.006525 & 27.662 & 28.0 & 4.23 & 14.8$\pm$1.0 \\
        &          &          & 12951 & 71.95    &  \\
        &          &          & 12953 & 31.76    & \\
        &          &          & 13246 & 45.02    & \\
        &          &          & 13247 & 34.08    & \\
        &          &          & 13255 & 43.34    & \\
NGC5846 & 15:06:29.253 & +01:36:20.29 &  7923 & 85.25    & E    & $\surd$  & 0.00491  & 20.867 & 21.1  &  4.24 & 21.0$\pm$1.3 \\
NGC6338 & 17:15:23.0 & +57:24:40.0 & 4194 & 44.52 & E5 &           & 0.027303 & 112.10 & 118.3 & 2.55 & 57.0$\pm$1.8 \\
NGC6482 & 17:51:48.833 & +23:04:18.88 & 3218 & 10.03 & E &            & 0.013129 & 55.091 & 56.5 & 8.04 & 3.2$\pm$0.1 \\
NGC6861 & 20:07:19.482 & -48:22:12.94 & 11752 & 88.89 & SA0 &        & 0.009437 & 39.826 & 40.6 & 4.94 & 40.7$\pm$2.0 \\
NGC7618 & 23:19:47.212 & +42:51:09.65 & 16014 & 121.00   & E &         & 0.017309 & 72.164 & 74.7  &  11.9 & 38.3$\pm$2.0 \\
UGC408  & 00:39:18.578 & +03:19:52.87 & 11389 & 93.80 & SAB & & 0.014723 & 61.628 & 63.5 & 2.80 & 980.0$\pm$30.9 \\
\hline
\end{tabular}
\end{table*}

\section{Sample and Data Reduction}\label{sec:2}
\subsection{Sample selection}

Our sample was selected from \citet{Babyk:17scal}  which includes 94 relatively nearby, early-type galaxies (BCGs, brightest group galaxies (BGGs), and early spiral galaxies) observed with the $Chandra$ X-ray Observatory.  The sample is intended to investigate X-ray scaling relations, structural properties, and dynamical properties of early-type galaxies over a large range in mass \citep{Main:17, Hogan:17, Hogan:17a, Pulido:17}. The challenge of assembling spatially resolved profiles of lower mass early-type systems concerns low count rates and relatively short exposure times. Previous studies focused on the $10-15$ brightest galaxies and groups (see, e.g., \citet{Werner:13}).  Recently, \citet{Werner:18new} presented the spatial analysis of 49 early-type galaxies. Here, we examine 40 early-type galaxies with a sufficient number of photon events to extract at least four concentric annular bins. The sample spans temperatures of $\sim$ 0.2--2.5 keV, X-ray luminosities of $L_{\rm X} \sim $ 10$^{37-42}$ erg/s, radio luminosities of $L_{\rm R} \sim$ 10$^{35-42}$ erg/s  corresponding to estimated jet powers of $P_{\rm jet} \sim$ 10$^{40-45}$ erg/s. Redshifts lie between $z \sim$ 0.001--0.032.

Interpreting trends based on archival analyses can be perilous owing to the poorly-defined selection function.  The targets tend to be relatively bright in X-rays and were selected for observation based on heterogeneous criteria.  This concern is lessened to some degree by cross-correlating the X-ray observations with molecular gas data taken from the ATLAS${}^{3D}$ catalog, which has a well-defined selection function. ATLAS${}^{3D}$ molecular gas masses and upper limits were measured with single dish IRAM observations \citep{Young:11}. The ATLAS${}^{3D}$ objects were then cross correlated with the early-type galaxies in \citep{Babyk:17scal, Babyk17nature} to form the sample presented here. The ATLAS${}^{3D}$ is a complete sample of early-type galaxies brighter than $M_K = -21.5$ within a distance of 42 Mpc \citep{Young:11}.  

Single dish CO measurements from \citet{Edge:01}, subsequently analyzed by \citet{Pulido:17}, form the cluster central galaxy sample that we compare to the early-type galaxy sample.  The Edge sample selected central galaxies in cooling cores with H$\alpha$ emission above L$_{H\alpha}$ = 10$^{39}$ erg/s. Thus the Edge selection differs from ATLAS${}^{3D}$, which will  limit the conclusions that can be drawn.  Nevertheless, the samples combined allow us to examine, for the first time, relationships between molecular gas and hot atmospheres over a wide range of halo mass. This is a significant step.  Other studies have focused on the relationship between molecular gas and the stellar content and their dynamics \citep{Emsellem:01, Emsellem:11, Young:08} or the relationship between atmospheric properties,  the stars, and the central black hole \citep{Ma:14}. This study extends our work on central galaxies in clusters, whose molecular gas is closely tied to their atmospheres \citep{Edge:02, McNamara:14, Russell:14, Russell:15, Russell:17a, David:14, David:17, Tremblay:16, Vantyghem:17, Vantyghem:18, Pulido:17},  to lower mass atmospheres and their parent halos. 

Table~\ref{tab1} lists the sample, including target name, coordinates, $Chandra$ observational ID, cleaned exposure time, morphology, galaxy classification, redshift, angular and luminosity distance, foreground hydrogen column density taken from \citet{Dickey:90}, and radio flux. Properties taken from NED\footnote{https://ned.ipac.caltech.edu/}, SIMBAD\footnote{http://simbad.u-strasbg.fr/} and HyperLEDA\footnote{Lyon-Meudon Extragalactic Database} databases. We found a slight discrepancy in morphological definition between NED and SIMBAD. For example, the NGC507 and NGC4382 are classified as SA0 galaxies in NED and S0 in SIMBAD. Our sample includes 11 BCGs. The angular and luminosity distances were defined using their redshift for the cosmology described above and, for some Virgo's galaxies, using the surface brightness fluctuations \citep{Mei:07, Cappellari:11}.

\subsection{Data processing}

Data reduction was done using {\sc ciao} version 4.8 and {\sc caldb} version 4.7.1 (see \citealt{Babyk:17scal, Hogan:17} for more details).  $Chandra$ observations were downloaded from HEASARC\footnote{http://heasarc.gsfc.nasa.gov/} archive.  Data processing of the event lists was performed with a custom-made pipeline {\sc xpipe} \citep{Fruscione:06}. The re-processing and re-screening of data by creating new bad pixel files and level 2 event files were achieved using the \texttt{chandra\_repro} task. The correction of time-dependent gain was applied using \texttt{acis\_process\_events} task. Background flares were flagged and removed using \texttt{lc\_clean} task provided by M. Markevitch. Periods of anomalously high backgrounds were excluded. Cosmic ray afterglows were removed using \texttt{acis\_detect\_afterglow} task.  Multiple observations were reprojected to the position of the observation with the longest exposure time. Each observation was processed with the appropriate blank-sky background file, which was normalized and reprojected to the corresponding position.  Each observation includes a 0.3 -- 6.0 keV energy band. For multiple observations, the images were summed and backgrounds were subtracted.  The \texttt{wavdetect} routine was applied to detect and remove point sources. 

\subsection{Spectral extraction and modeling}


Source and background spectra were extracted individually for each observation. The ancillary response and response matrix files were generated using {\sc mkwarf} and {\sc mkacisrmf} tools. The spectra were grouped with at least 20 counts per energy bin using the {\sc ftools} task {\sc grppha}.  Corrections for chip gaps and the area lost to point sources were corrected using exposure maps.
\begin{figure*}    
\centering
\includegraphics[width=0.33\textwidth]{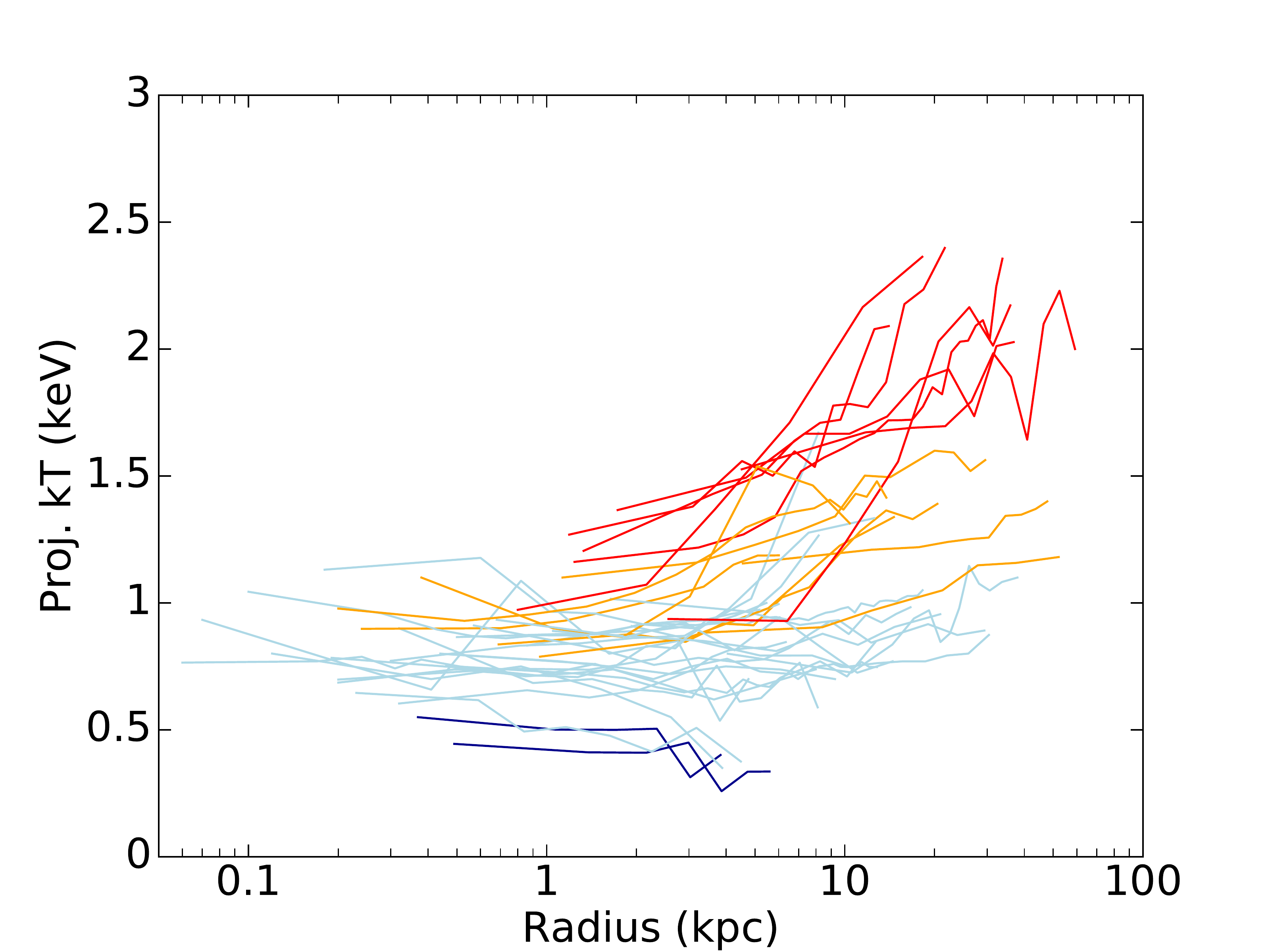}
\includegraphics[width=0.33\textwidth]{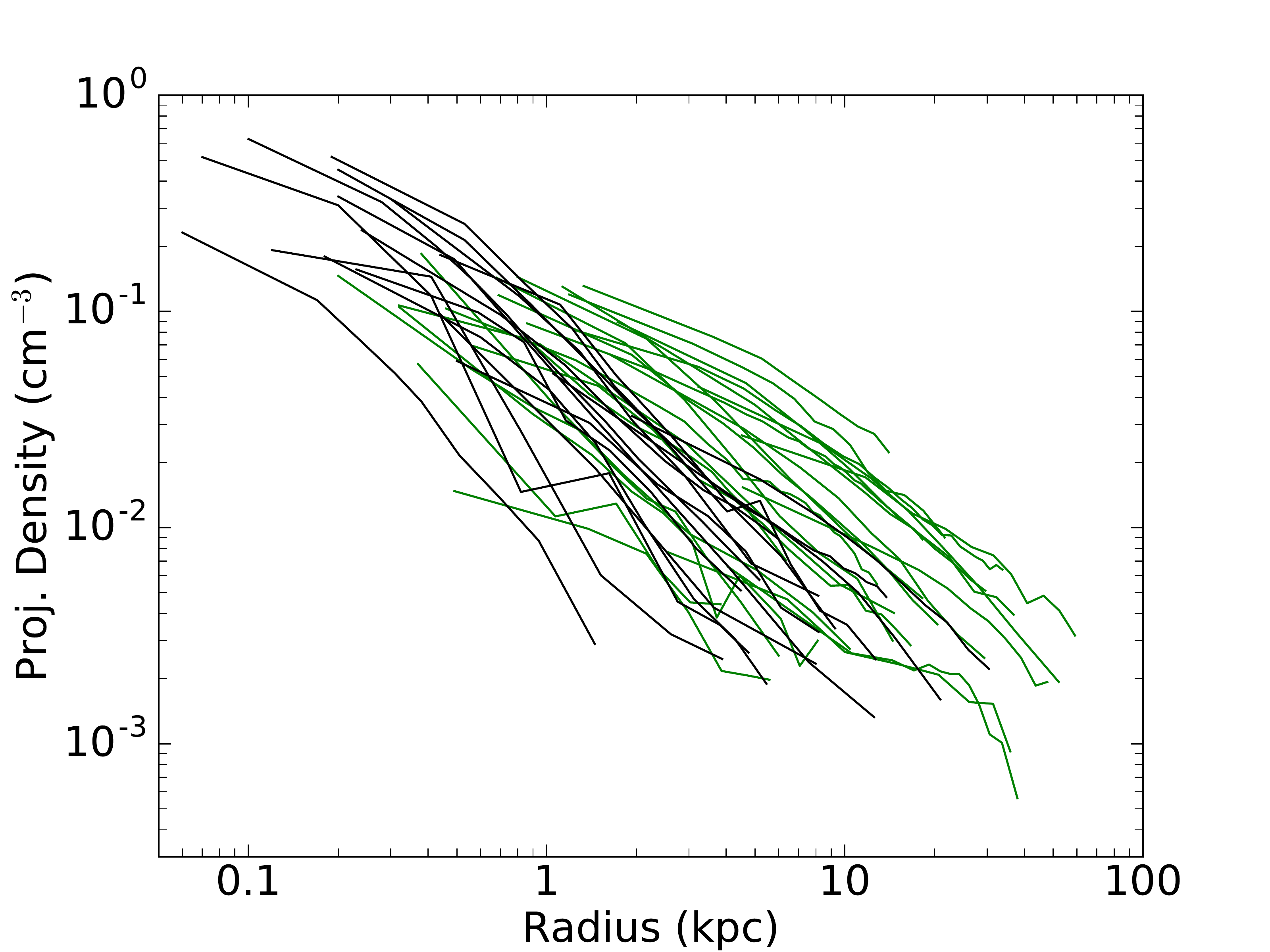}
\includegraphics[width=0.33\textwidth]{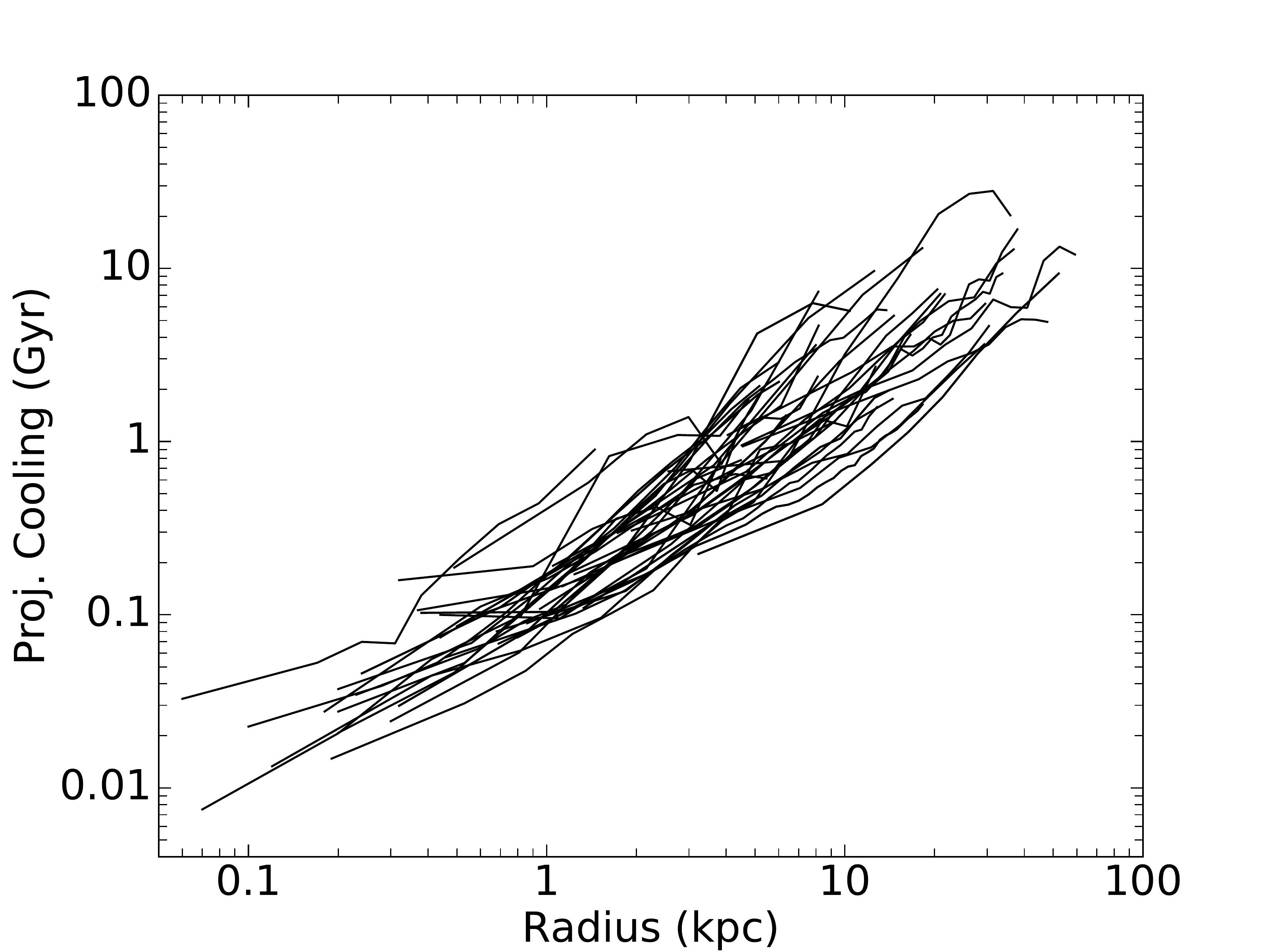}
\includegraphics[width=0.33\textwidth]{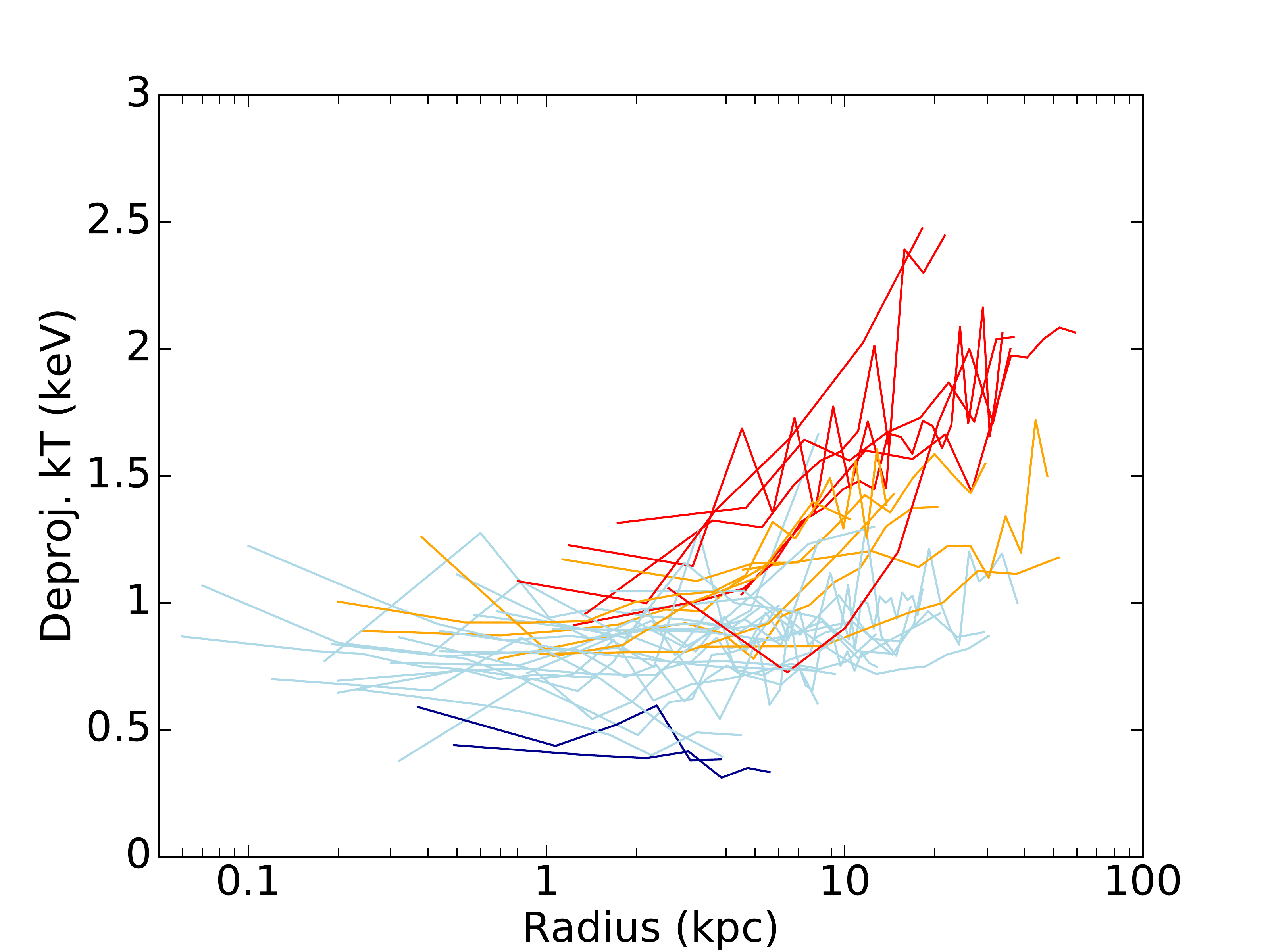}
\includegraphics[width=0.33\textwidth]{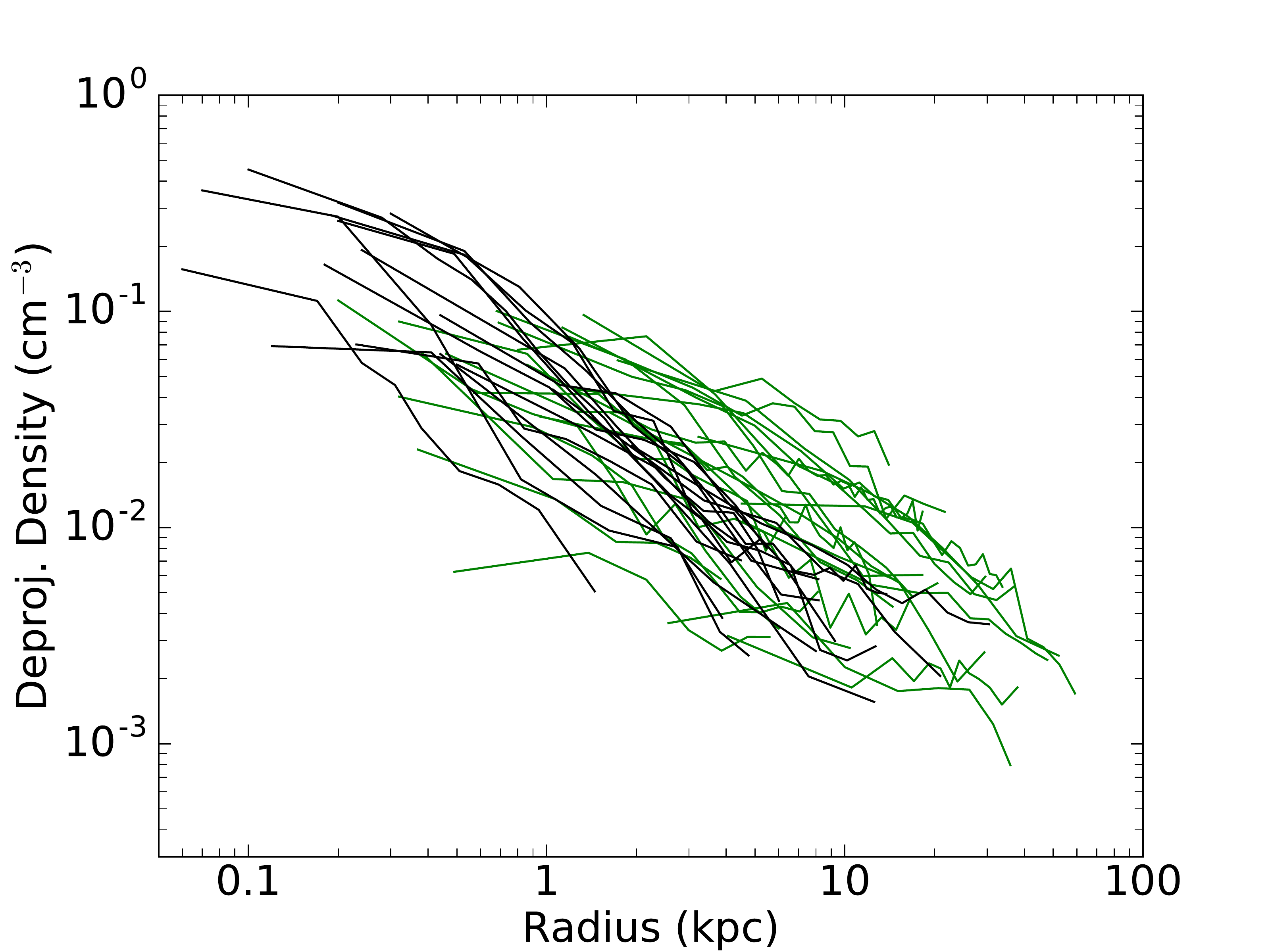}
\includegraphics[width=0.33\textwidth]{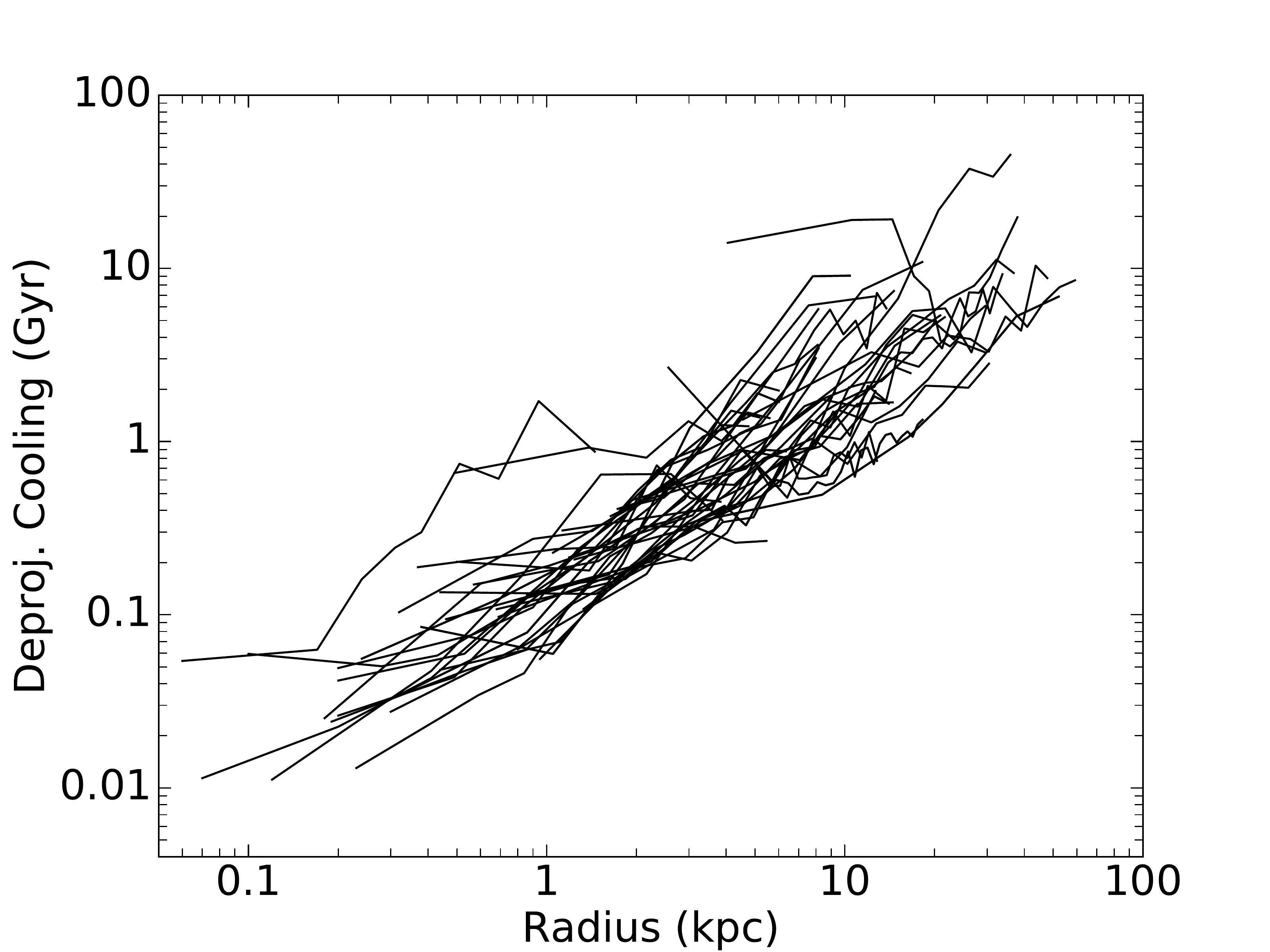}
\caption{Projected (top) and deprojected (bottom) temperature, density and cooling time profiles (from left to right) for entire sample of low-mass systems. For clarity we plot temperature profiles using four colors. Dark blue temperature curves show profiles with average temperature in the range of 0.2-0.5 keV, light blue 0.5-1.0 keV, orange 1.0-1.5 keV, while red 1.5-2.5 keV. Density profiles are indicated by shape. Those characterized by power laws of the form $\beta = 1.5-2.0$ are indicated in black. Density profiles with cores, $\beta = 1.0-1.5$, are indicated in green. We removed error bars for clarity.}
\label{fig_tdprof}
\end{figure*}

Spectral analysis was done with {\sc xspec} version 12.9.1 (\citealt{Arnaud:96}). The spectral fitting included an absorbed, single-temperature, multi-component model {\sc phabs*(apec+po+mekal+po)} to correct for unresolved, low-mass X-ray binaries (LMXBs), active binaries (AB), and cataclysmic variables (CV).  The {\sc apec} component \citep{Smith:01} describes the emission of hot gas, while the first {\sc po} and set of {\sc mekal+po} models \citep{Mewe:86, Kaastra:93, Liedahl:95} fit the possible contribution of the X-ray emission of LMXBs and ABs+CVs sources, respectively. The hydrogen column density, $N_{\rm H}$, was fixed as given in Tab.~\ref{tab1}.  

Metallicity in the {\sc apec} model was fixed at 0.5$\rm Z_{\odot}$. We also perform the spectral fitting with free $N_{\rm H}$. We find that the best-fit $N_{\rm H}$ for about 40\% our sample is higher by a factor of 1.5-2 compared to those obtained by \citet{Dickey:90} and is in agreement with more recent measurements of \citet{Kalberla:05}. However, the best-fit temperature and normalization are unaffected by this discrepancy. The parameters of additional spectral components that were added to represent the emission of LMXBs and stellar sources emission were fixed (see, e.g., \citet{Babyk:17scal} for more details of presented spectral model and, e.g., \citet{Boroson:10, Wong:14} for other multi-component spectral models). The spectra were fitted in the 0.3 -- 6.0 keV energy range which provides an optimal ratio of the galaxy and background flux for $Chandra$ observations. Multiple spectra for the same annulus were fitted simultaneously. Uncertainties were determined using {\sc xspec} \texttt{error}  quoted for 1$\sigma$ confidence. Quoted uncertainties are statistical.

\section{Galaxy properties}\label{sec:3}
Here we present profiles of temperature, density, cooling, mass and free-fall time for the range of radii $\sim$ 0.1--50.0 kpc. Analysis of entropy profiles is presented in \citet{Babyk17nature}. The multi-component spectral model described above was applied, and the output temperature and normalization parameter of {\sc apec} model were  extracted. The X-ray emission from each extraction annulus includes emission from other hotter annuli at higher altitudes. Therefore, the emission must be ``deprojected'' to isolate the emission in the desired annulus from emission from overlying layers.  For this the model-independent {\sc DSDEPROJ} routine \citep{Russell:08} was adopted. The deprojected spectra were then fitted using the multi-component model mentioned above in the same way as projected spectra.

\subsection{Temperature, density, and cooling time profiles}
On the left-hand panels of Fig.~\ref{fig_tdprof} we plot  temperature profiles before (top) and after (bottom) deprojection. For clarity we plot profiles using four colors. Dark blue lines show profiles with average temperature in the range of 0.2-0.5 keV, light blue 0.5-1.0 keV, orange 1.0-1.5 keV, while red 1.5-2.5 keV. Error bars are omitted for clarity. The projected and deprojected temperature profiles show significant variations. Galaxies with average temperatures below 1.5 keV tend to rise in temperature toward their centers. This may be due to shock activity, which raises the entropy of the central atmospheres of ellipticals and some BCGs \citep{Werner:12}. Flat temperature profiles are seen in others. The red temperature profiles indicate systems with sharp temperature increases (by a factor 1.5 and higher within just several kpc) beyond $\sim$ 10 kpc. Such a quick rise is probably associated with the hot intracluster medium since this rise is only observed in BCGs. 

Density profiles were constructed using the spectral temperature and normalization ($N$) parameters. The projected and deprojected electron number density profiles were derived as
\begin{equation}
n_{\rm e}~=~D_{\rm A}(1+z)10^{7}~\sqrt[]{\frac{4\pi~N~1.2}{V}},
\end{equation}
where the factor of 1.2 was defined using the ionization ratio $n_{\rm e}$/$n_{\rm p}$, $V$ is the volume of concentric annular region in cm$^3$, $D_{\rm A}$ is the angular diameter distance, and  $N$ is the normalization, which is proportional to the integrated emission measure. Density profiles follow $n_{\rm e} \propto r^{-\beta}$, with $\beta \approx$ 1.0-2.0. The middle panels of Fig.~\ref{fig_tdprof} show the radial distribution of projected and deprojected electron densities.  Density profiles are indicated by shape.  Those characterized by power laws of the form $\beta = 1.5-2.0$ are indicated in black.  Profiles with cores, $\beta = 1.0-1.5$, are indicated in green. The galaxies with X-ray power law density profiles (black lines) correspond to lower temperature atmospheres with $T = 0.3-1.0$ keV.  Core profiles in green are associated with hotter atmospheres $T \gtrsim$ 1 keV. The central projected densities are $10-30\%$ higher than deprojected profiles. 

\begin{figure*}[t!]    
\centering
\includegraphics[width=0.33\textwidth]{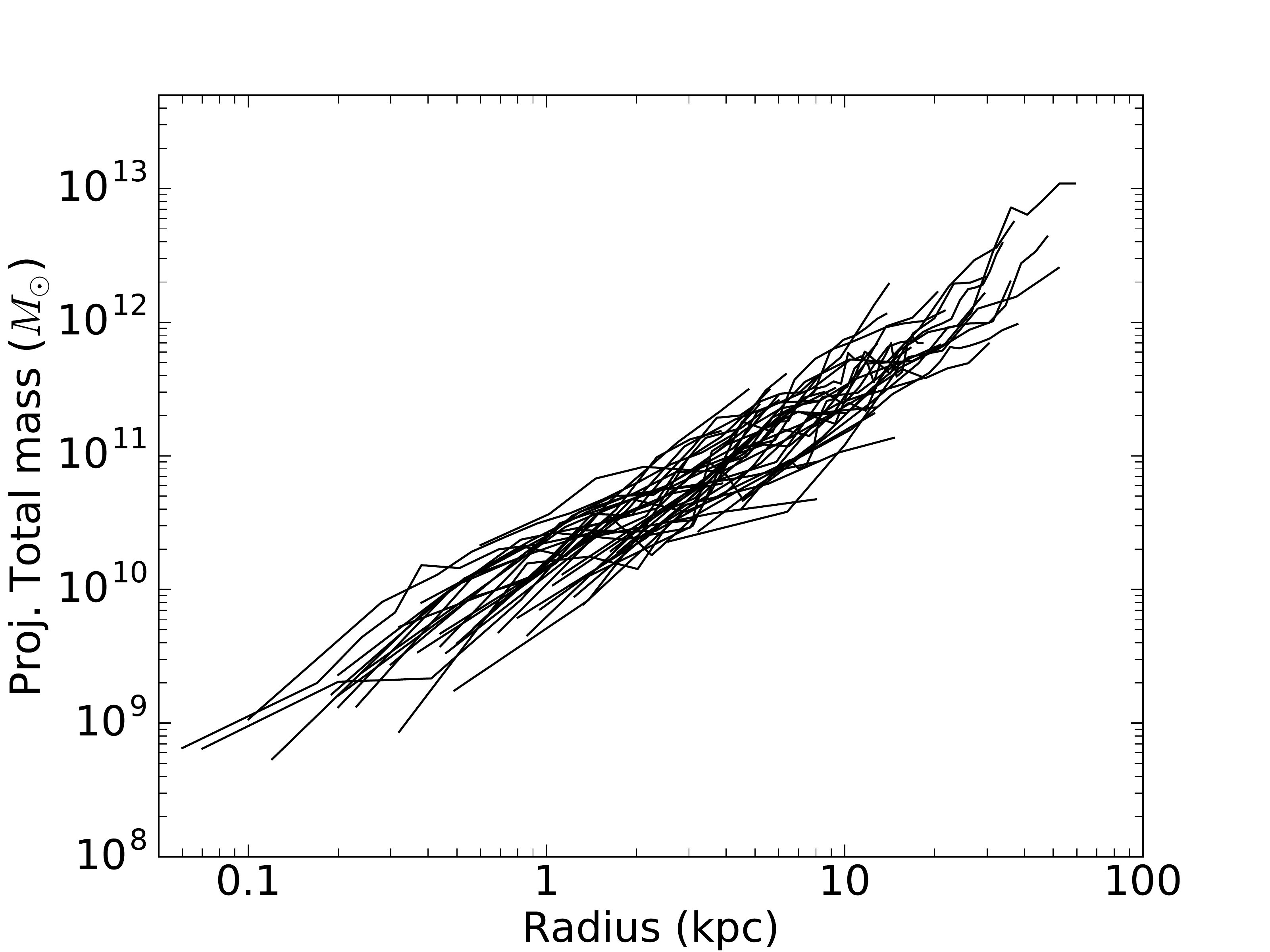}
\includegraphics[width=0.33\textwidth]{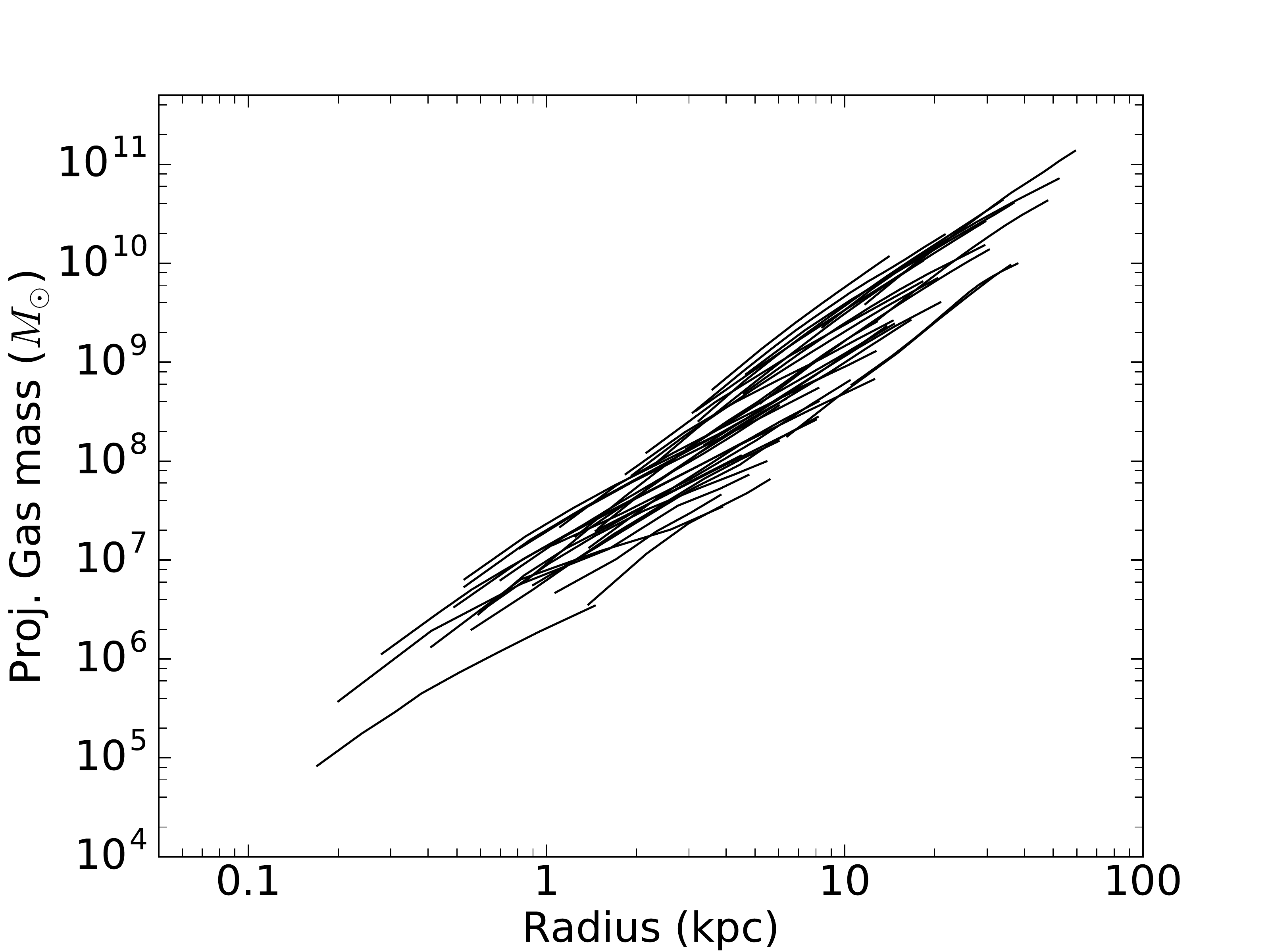}
\includegraphics[width=0.33\textwidth]{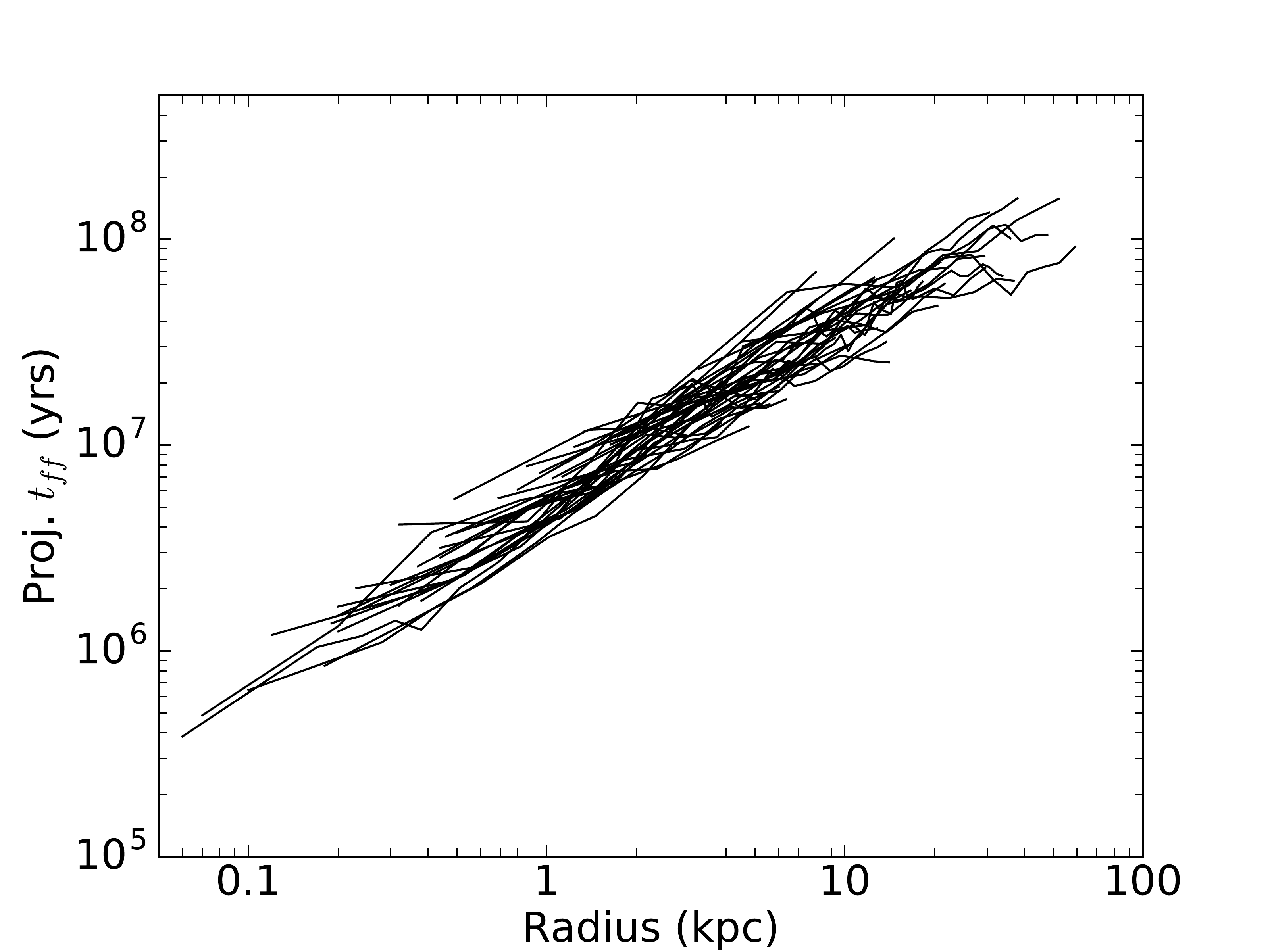}
\includegraphics[width=0.33\textwidth]{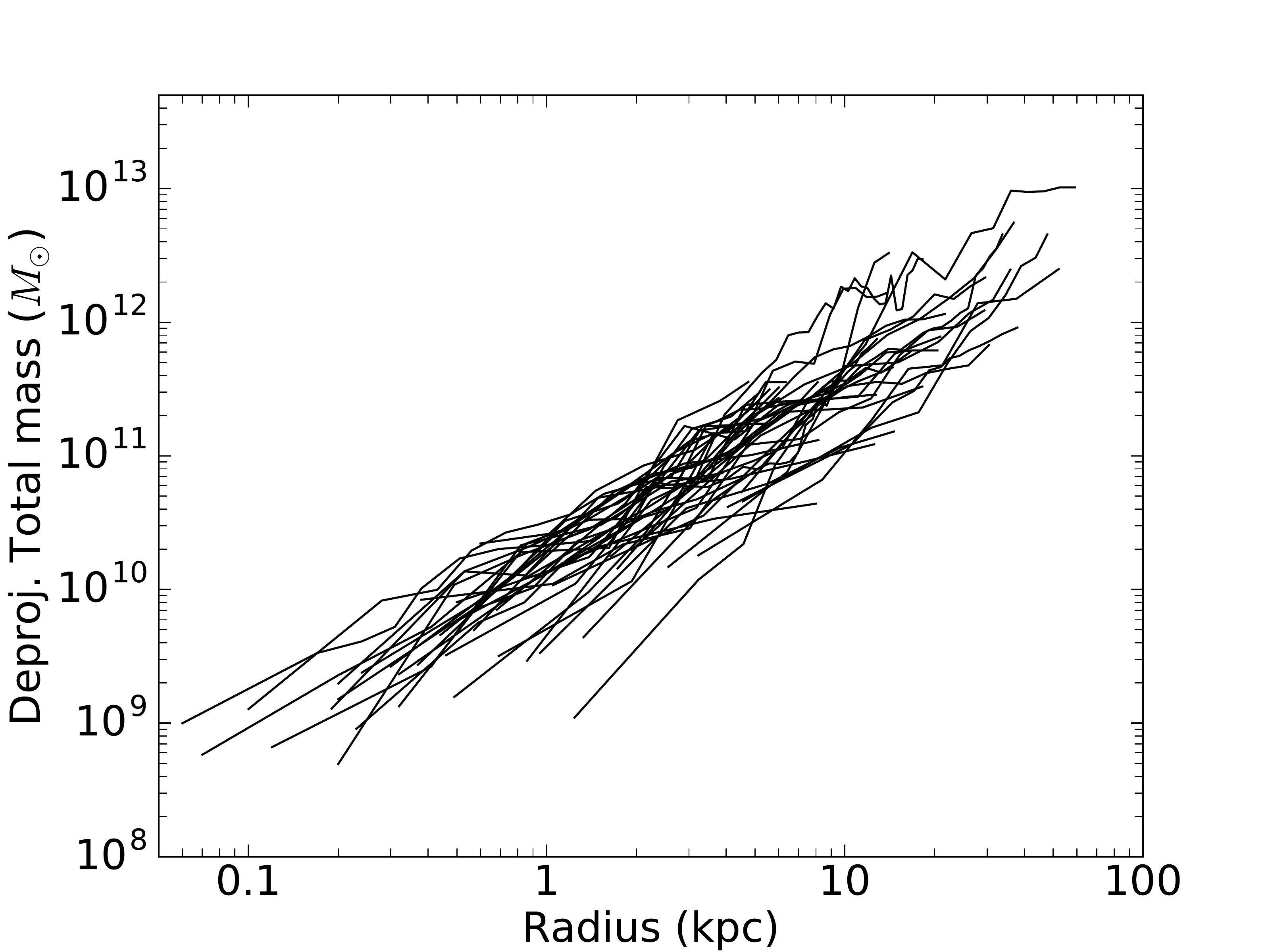}
\includegraphics[width=0.33\textwidth]{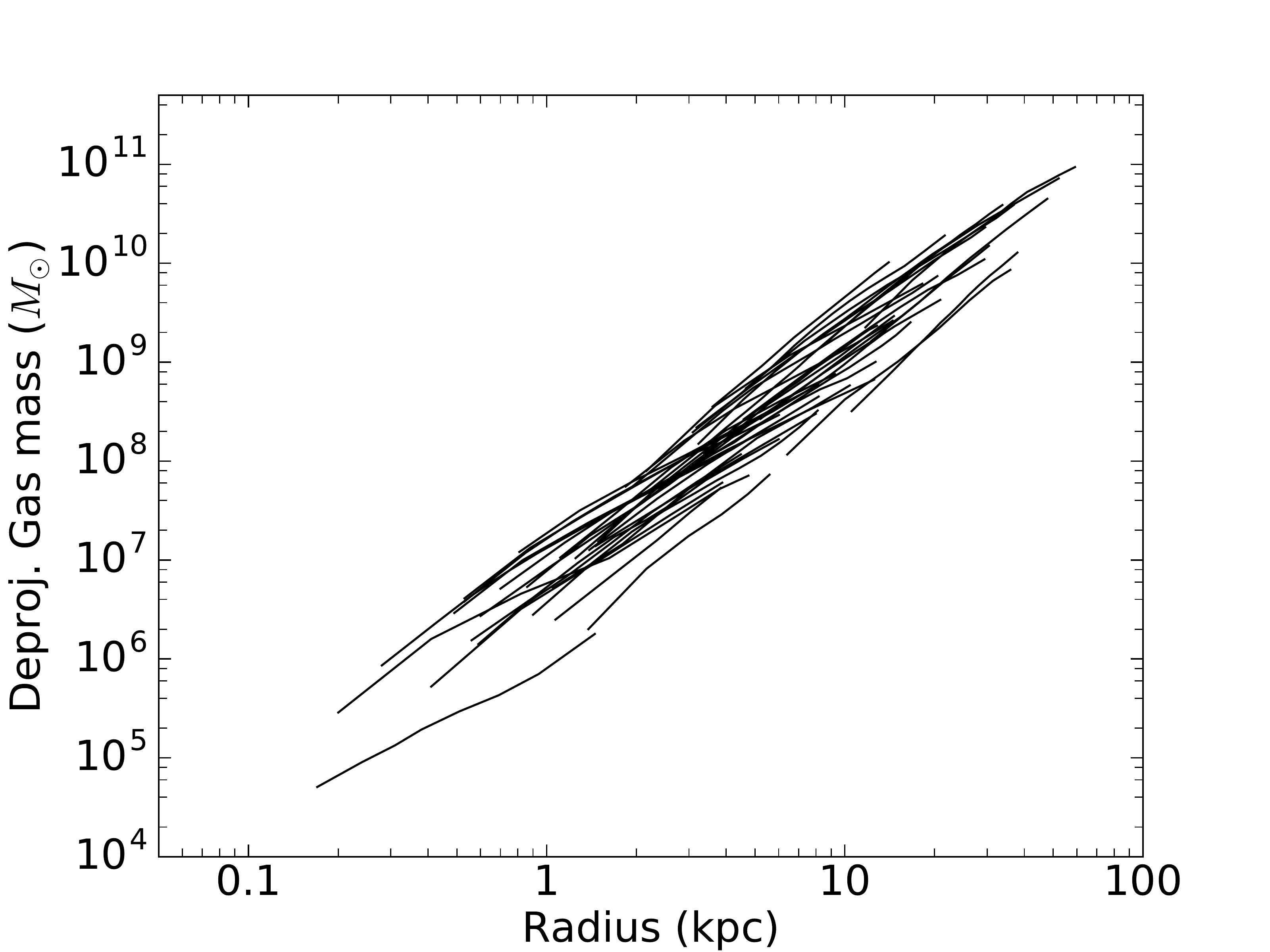}
\includegraphics[width=0.33\textwidth]{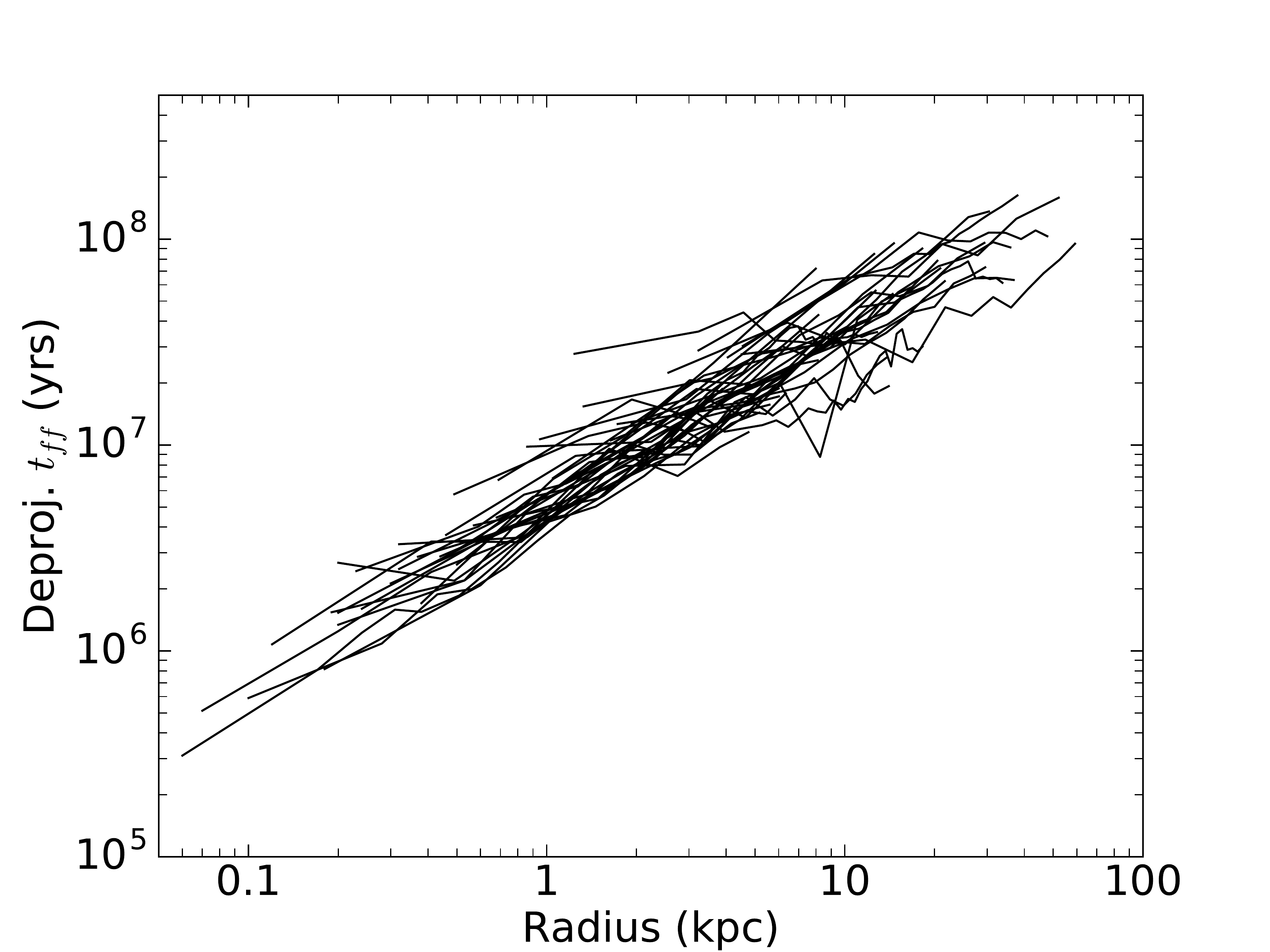}
\caption{Projected (top) and deprojected (bottom) total and gas mass profiles with the free-fall time profiles for entire sample of low-mass systems. We removed error bars for clarity.}
\label{fig_cfprof}
\end{figure*}

The hot atmosphere radiates its thermal energy on the cooling timescale given by  
\begin{equation}
t_{\rm c} = \frac{3P}{2n_{e\rm }n_{\rm H}\Lambda(Z,T)} \approx \frac{3PV}{2L_X}.
\end{equation}
This is the thermal energy of the gas, $E_{thermal} = PV$ divided by the energy lost per unit volume. Here $\Lambda(Z,T)$ is the cooling function that depends on metallicity and temperature. $P = 2 n_{\rm e} k_B T $ is the pressure and $L_{\rm X}$ is the X-ray luminosity. First, we extracted X-ray flux of the galaxies by integrating the multi-component model between the 0.1 -- 100 keV energy band. Then, an estimated X-ray flux was used to derive the bolometric X-ray luminosity as, $L_{\rm X} = 4\pi D_{\rm L}^{2}\times{\rm flux}$. The projected and deprojected cooling time profiles are shown on the right-hand panels of Fig.~\ref{fig_tdprof}. The cooling time of entire sample lies below $t_{\rm c} \lesssim$ 10$^9$ yr at $\lesssim$ 10 kpc. The largest cooling time is 14.04$\pm$2.31 Gyr within the innermost region ($\sim$ 4.05 kpc) in UGC408 galaxy, while the lowest one is about 0.01 Gyr at 0.1 kpc in IC1459 and NGC1332 targets. 

\subsection{Mass and free-fall time profiles}
\subsubsection{The gas and total mass}
We use assumptions of hydrostatic equilibrium and spherical symmetry of hot gas in the gravitational potential (see, e.g., \citealt{Navarro:10, Babyk:13, Babyk:13oap, Babyk:14, Babyk_book:15, Babyk:16}) to calculate the total mass profiles. The derived temperature and density profiles were used to estimate projected and deprojected total mass as
\begin{equation}
M_{T}(r) = -\frac{k_BT(r)r}{G \mu m_{p}} \left( \frac{d \ln n_{e} }{d \ln r} + \frac{d \ln T}{d \ln r}\right),
\label{mtr}
\end{equation}
where $\mu$ = 0.62 is the mean particle mass in units of proton mass, $k_B$ is the Boltzmann constant, $G$ is the gravitational constant, and $m_{\rm p}$ is the proton mass. The projected and deprojected total mass profiles for the entire sample of low-mass systems are given in the left-hand panels of Fig.~\ref{fig_cfprof}. The two sets of mass profiles are consistent with one another and they show a scatter of $\sim$ 20\% in mass at a fixed radius. The total mass spans the ranges $\sim$ 10$^{9-13}\rm M_{\odot}$ and $\sim$ 0.1--50.0 kpc in radius. 

We derive gas mass profiles by integrating the gas density over the radius. The results of this integration are shown in the middle panels of Fig.~\ref{fig_cfprof} for both projected and deprojected profiles. The gas mass profiles are similar, showing a scatter of only $\sim$ 15\%. The gas mass spans 6 decades, $\sim$ 10$^{5-11}\rm M_{\odot}$. The gas fractions are small, about 1\% at 10 kpc.

\subsubsection{Free-fall time}
One aim of this paper is to understand the effect of mass in regulating the balance between cooling and heating in galaxy and cluster cores. We examine the role of mass in two ways.  First, by evaluating correlations between molecular gas mass and the mass of various components of early-type galaxies, such as atmospheric gas mass, stellar mass, and total mass within a given radius.  Secondly we look at acceleration, which is related to mass.  The free-fall time profiles are useful because they are relevant to thermally unstable cooling of hot atmospheres.  We derive free-fall time profiles using the total mass profiles to calculate acceleration, $g = (G M)/r^2$, as 
\begin{equation}
t_{\rm ff}(r) = \sqrt{2r/g}.
\label{eqtff}
\end{equation}
Free-fall time profiles of the form (Eq.~\ref{eqtff}) can be directly compared to previous results in clusters. The right-hand plots in Fig.~\ref{fig_cfprof} show the projected and deprojected free-fall time profiles. They are consistent with each other showing only $\sim$ 14\% variation at 10 kpc. Free-fall time profiles have been measured for a large number of central galaxies in clusters \citep{Hogan:17, Hogan:17a, Main:17, Pulido:17}. These studies found, as we do here, a small variance in the free-fall time profiles. Furthermore,  their value at 10 kpc, $\simeq 5\times 10^7$ yr, found in cluster central galaxies are close to those found here for early-type galaxies. This demonstrates that the mass profiles of BCGs and early-type galaxies are similar in their inner few tens of kpc.

\section{Thermally Unstable Atmospheric Cooling}\label{sec:4}
Hot atmospheres are expected to become thermally unstable to linear density perturbations, when the ratio of the cooling time to free-fall time, \tctff\, falls below unity \citep{Nulsen:86, Pizzolato:05, McCourt:12}. Hydrodynamical simulations of three-dimensional atmospheres apparently show that this instability criterion threshold may rise well above unity.  Studies have claimed that thermally unstable cooling ensues from linear perturbations when \tctff\ falls below 10, fueling a self-regulating feedback loop \citep{Sharma:12, McCourt:12, Gaspari:13, Li:15}. However, later simulations by \citet{Sharma:16} do not confirm the theoretical basis for the \tctff\ $<$ 10 criterion. Furthermore, using a small number of systems, \citet{McNamara:16} showed that the observed ratio of \tctff\ is driven by the numerator, $t_{\rm c}$, while the denominator, $t_{\rm ff}$ adds noise (see Discussion section for more details). Since then, larger samples have been analyzed, paying careful attention to systematic biases, spanning a large range of halo mass.  Analyses of \tctff\  profiles of clusters and their BCGs performed by \citet{Hogan:17a} and \citet{Pulido:17} showed no evidence that \tctff\ falls significantly below 10.  They found  that \tctff\ lies between 10 and 30 in systems with star formation and molecular clouds, with no indication that lower values of \tctff\ correlated with higher star formation rates or molecular gas masses.  Consistent with \citet{Voit:15, Voit:19}, they found a floor at \tctff\ $\sim$ 10 rather than a threshold.  While this floor may well be physically significant, they found that the range of \tctff\ values can be explained as an observational selection effect, raising uncertainty in its interpretation.

\begin{figure}    
\includegraphics[width=0.49\textwidth]{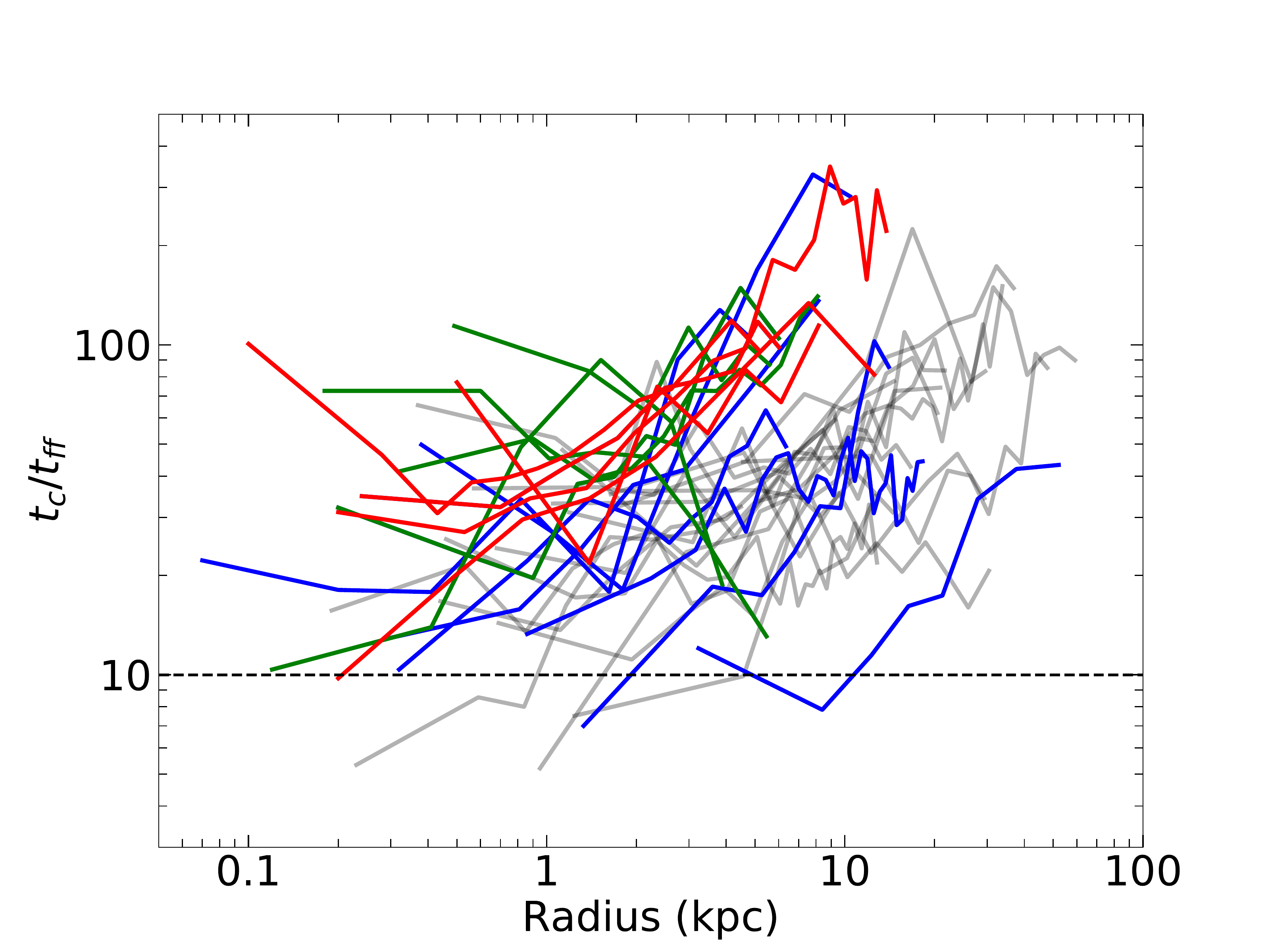}
\caption{Deprojected \tctff\ profiles for the entire sample of ETGs. Error bars are deleted for clarity. Blue profiles are objects with CO detection. Red and green profiles are upper limits.}
\label{fig_tctfprof}
\end{figure}

In Fig.~\ref{fig_tctfprof} we plot profiles of the deprojected ratio \tctff. We find that the minima all lie between $10-30$.  While a few values dip below 10, these departures are consistent with measurement noise. Plotted in blue are the \tctff\ profiles for systems with CO detections. Plotted in green and red are CO upper limits below 10$^8$ and 10$^7$ solar mass respectively. The systems with CO detections do not segregate from those with CO upper limits. However, those that contain detectable levels of CO preferentially show lower cooling times, a topic discussed below. 

\section{Molecular gas distribution}\label{sec_molec}
\begin{table*}
\caption{Early-type galaxies of ATLAS${}^{3D}$ sample.}\label{tab3}
\centering
\begin{tabular}{lcccccccccccccc}
\hline
 && \\
Name      & R.A. & Decl. & ObsIDs & Exposure & Type & $z$ & $D_{\rm A}$ &  $D_{\rm L}$  &  $N_{\rm H}$ & kT & $\log(M_{\rm mol})$ & 1.4 GHz Flux\\
          &  (J2000)   & (J2000)&        &    ks    &     &      &   Mpc  &  Mpc   &  10$^{20}$ cm$^{2}$ & keV & $\rm M_{\odot}$ & mJy\\
          & (2) & (3) & (4) & (5) & (6) & (7) &(8) & (9) & (10) & (11) & (12) & (13)\\
&& \\
\hline
&&\\
IC1024   & 14:31:27.221 & +03:00:32.78 & 14901 & 20.07  & S0  & 0.004933 & 20.96  & 21.2 & 2.55 & 0.46$\pm$0.11 & 8.61$\pm$0.02 & 24.7$\pm$1.2\\
NGC1266  & 17:45:35.288 & -46:05:23.71 & 19498 & 80.06  & Sb  & 0.007238 & 30.65  & 31.1 & 5.18 & 0.80$\pm$0.03 & 7.58$\pm$0.01 & 115.6$\pm$3.5\\
NGC2768  & 09:11:37.504  & +60:02:13.95 & 9528  & 65.46  & E6  & 0.004513 & 19.19  & 19.4 & 3.87 & 0.35$\pm$0.02 & 7.64$\pm$0.07 & 14.5$\pm$0.6\\
NGC3245  & 10:27:18.385  & +28:30:26.63  & 2926  & 9.76   & Sab & 0.00423  & 13.827 & 13.9 & 2.08 & 0.55$\pm$0.11 & 7.27$\pm$0.12 & 6.7$\pm$0.5\\
NGC3599  & 11:15:26.949 & +18:06:37.43 & 9556  & 20.16  & S0  & 0.002799 & 11.93  & 12.0 & 1.42 & 0.36$\pm$0.05 & 7.36$\pm$0.08 & 2.3$\pm$0.4\\
NGC3607  & 11:16:54.657 & +18:03:06.51 & 2073  & 39.00  & E   & 0.003142 & 13.39  & 13.5 & 1.48 & 0.59$\pm$0.11 & 7.42$\pm$0.05 & 6.9$\pm$0.4\\
NGC3619  & 11:19:21.621 & +57:45:27.66 & 19320 & 9.97   & S0  & 0.005204 & 22.107 & 22.3 & 74.7 & 0.36$\pm$0.14 & 8.28$\pm$0.05 & 5.6$\pm$0.5\\
NGC3665  & 11:24:43.630 & +38:45:46.05 & 3222  & 18.19  & S0  & 0.006901 & 29.23  & 29.6 & 2.06 & 0.30$\pm$0.05 & 8.91$\pm$0.02 & 112.2$\pm$3.7\\
NGC4036  & 12:01:26.891 & +61:53:44.52 & 6783  & 15.13  & Sa  & 0.00462  & 19.64  & 19.8 & 1.89 & 0.46$\pm$0.07 & 8.13$\pm$0.04 & 11.6$\pm$0.5\\
NGC4203  & 12:15:05.054 & +33:11:50.40  & 10535 & 42.18  & Sa  & 0.003623 & 15.429 & 15.5 & 1.19 & 0.28$\pm$0.03 & 7.39$\pm$0.05 & 6.1$\pm$0.5\\
NGC4283  & 12:20:20.769 & +29:18:39.16  & 7081  & 112.14 & E   & 0.003522 & 15.0   & 15.1 & 1.77 & 0.35$\pm$0.04 & 7.10$\pm$0.09 & 385.0$\pm$11.6\\
NGC4459  & 12:29:00.026 & +13:58:42.89  & 11784 & 30.18  & S0  & 0.003976 & 16.92  & 17.1 & 2.67 & 0.54$\pm$0.08 & 8.24$\pm$0.02 & 2.7$\pm$0.6\\
NGC4476  & 12:29:59.081 & +12:20:55.21  & 3717  & 20.83  & S0  & 0.006565 & 27.83  & 28.2 & 2.54 & 0.35$\pm$0.05 & 8.05$\pm$0.04 & 245.6$\pm$8.7\\
NGC4477  & 12:30:02.172 & +13:38:11.19  & 9527  & 38.18  & S0  & 0.004463 & 18.98  & 19.2 & 2.63 & 0.34$\pm$0.02 & 7.54$\pm$0.06 & 6.2$\pm$0.5\\
NGC4526  & 12:34:03.029 & +07:41:56.90 & 3925  & 44.11  & S0  & 0.002058 & 16.5   & 17.5 & 1.65 & 0.37$\pm$0.02 & 8.59$\pm$0.01 & 12.0$\pm$0.5\\
NGC4596  & 12:39:55.954 & +10:10:34.18 & 11785 & 31.38  & Sa  & 0.006311 & 27.76  & 27.1 & 1.98 & 0.27$\pm$0.05 & 7.31$\pm$0.09 & 3.4$\pm$0.6\\
NGC4710  & 12:49:38.958 & +15:09:55.76 & 9512  & 30.16  & S0  & 0.003676 & 15.65  & 15.8 & 2.15 & 0.64$\pm$0.06 & 8.72$\pm$0.01 & 18.7$\pm$1.0\\
NGC5866  & 15:06:29.561 & +55:45:47.91 & 2879  & 27.43  & Sab & 0.002518 & 14.9   & 15.2 & 1.45 & 0.41$\pm$0.08 & 8.47$\pm$0.01 & 21.8$\pm$1.1\\
NGC7465  & 23:02:00.952 & +15:57:53.55 & 14904 & 30.06  & Sb  & 0.006538 & 27.71  & 28.1 & 6.03 & 0.45$\pm$0.10 & 8.79$\pm$0.02 & 19.1$\pm$1.1\\
\hline
\end{tabular}
\end{table*}

The primary aim of this study is to understand the origin of molecular gas in the cores of low- and high-mass systems.  Archival ALMA data are available for five galaxies in our sample, which we analyze here. We use previously published cold gas mass measurements for clusters and early-type galaxies to explore possible relationships between atmospheric properties and molecular gas, as found in earlier studies of cluster central galaxies  \citep{Cavagnolo:08, Rafferty:08, Voit:15a, Pulido:17}. As most CO measurements of early-type galaxies have yielded upper limits, we use survival statistic to examine our relations. We also use a $\beta$-model \citep{Cavaliere:78, Babyk:12lviv, BabykDel:14} fitting to calculate density and gas mass profiles of those galaxies with known masses of cold gas but with X-ray data too poor to construct spatially resolved gas mass profiles. The objects  whose X-ray masses were estimated using $\beta$-model are listed in Table~\ref{tab3}. We follow \citet{Babyk:17scal} to fit the X-ray surface brightness profiles and to estimate the total and gas mass profiles.

\subsection{ALMA data reduction}
We analyzed archival ALMA data for five galaxies from our sample. IC1459, NGC1332, and NGC6861 were observed in ALMA band 6 to cover the CO(2-1) line emission, while NGC1407 and NGC4696 were observed in band 3 to cover CO(1-0) line emission. A description of observational parameters are listed in Table~\ref{tbl-obsalma}. The data were processed in \textsc{casa} version 4.7.2 using the ALMA pipeline reduction scripts. We applied continuum phase self-calibration to each dataset which improved the image quality. The continuum for each galaxy was imaged using natural weighting. To isolate the molecular line emission, the continuum was subtracted using \texttt{uvcontsub}. The continuum subtracted measurement sets were imaged using natural weighting for better sensitivity with 20 km s$^{-1}$ channels. Integrated line emission maps were produced by integrating over all the channels containing line emission.

\begin{table*}
\scriptsize
\centering
\caption{ALMA Observational Parameters and Results\label{tbl-obsalma}}
\begin{tabular}{lcccccccccc}
\hline
 \multirow{2}{*}{\phantom{m}Galaxy} & \multicolumn{1}{c}{Obs.} & \multirow{2}{*}{$t_{\rm obs}$} & \multicolumn{1}{c}{Synthesized} & \colhead{Frequency} & \multirow{2}{*}{Bandwidth} & \multicolumn{1}{c}{Velocity} & \multirow{2}{*}{PI} & S$_{CO(2-1)}$ & S$_{CO(1-0)}$ & $M_{\rm mol}$ \\
  & \multicolumn{1}{c}{Date} &  & \multicolumn{1}{c}{beam} & \multicolumn{1}{c}{Range} &  & \multicolumn{1}{c}{Resolution} & & & & \\
  &  & \multicolumn{1}{c}{(min)} & $(''\times'')$ & \colhead{(GHz)} & \colhead{(GHz)} & \colhead{(km/s)} &  & \colhead{(Jy km/s)} & \colhead{(Jy km/s)} & \colhead{($10^{7}$ $\rm M_\odot$)}\\
 \hline
 IC1459 & 2016-04-11 & 11.6 & $1.04\times0.80$ & $225.6-244.5$ & 5.9 & 2.56 & Prandoni Isabella &  <0.35 & - & <0.08\\
 NGC1332 & 2014-09-01 & 22.3 & $0.32\times0.24$ & $226.4-246.1$ & 5.9 & 1.28 & Barth Aaron & $38.72\pm 0.31$ & - & $6.04\pm 0.05$\\
 NGC1407 & 2016-05-03 & 15.0 & $1.50\times1.08$ & $112.1-115.3$ & 4.9 & 81.14 & Hodges-Kluck Edmund & - & <1.56 & <1.06\\
 NGC4696 & 2016-01-27 & 77.7 & $2.29\times1.74$ & $99.3-115.0$ & 7.6 & 5.22 & Hamer Stephen & - & 3.07$\pm$0.26 & 5.79$\pm$0.41\\
 NGC6861 & 2014-09-01 & 23.9 & $0.37\times0.28$ & $225.5-245.1$ & 5.9 & 1.28 & Barth Aaron & $89.91 \pm 1.62$ & - & 48.72$\pm$0.87\\ 
 \hline
 \end{tabular}
 \end{table*}

For NGC4696, the spectrum was fitted with a single Gaussian profile to get line intensity. For NGC1332 and NGC6861, spectra were integrated numerically and the errors in the line intensity ($\sigma_{I}$) were estimated using the formula given in \citet{Young:11} as 
\begin{equation}
\sigma_{I}^{2} = (\Delta \nu)^{2} \sigma^{2} N_{1} \bigg(1+\frac{N_{1}}{N_{b}}\bigg),
\end{equation}
where $\Delta \nu$ is the velocity channel width, $\sigma$ is the rms noise per channel, $N_{1}$ is the number of channels containing line emission and $N_{b}$ is the number of channels used to measure the baseline. For sources in which we do not detect any line emission, 3$\sigma$ upper limits on integrated line intensities were estimated by extracting the spectrum within 2 kpc $\times$ 2 kpc box centered at the optical centroid of the galaxies and following \citet{McNamara:94}
\begin{equation}
S_{\rm CO} \Delta \nu = \frac{3 \sigma_{\rm ch} \Delta V}{\sqrt{\Delta V/\Delta V_{\rm ch}}}\, {\rm Jy\,km\,s^{-1}},
\end{equation}
where $\sigma_{\rm ch}$ is the rms noise per channel in units of Jy, $\Delta V$ is the expected FWHM of the line, which we assume to be 200 km s$^{-1}$, and $\Delta V$ is the velocity channel width.

The molecular gas mass is calculated using integrated CO(1--0) line intensity as given in \citet{Solomon:05} and \citet{Bolatto:13}
\begin{equation}
\begin{multlined}
M_{\rm mol} = 1.05 \times 10^{4} \Bigg(\frac{X_{\rm CO}}{2 \times 10^{20} \frac{\rm cm^{-2}}{\rm K\, km\, s^{-1}}}\Bigg) \bigg(\frac{1}{1+z}\bigg) \times \\ \times \bigg(\frac{S_{\rm CO} \Delta \nu}{\rm Jy\, km\, s^{-1}}\bigg)
\bigg(\frac{D_{\rm L}}{\rm Mpc}\bigg)^{2}\, {\rm M_{\odot}},
\end{multlined}
\end{equation}
where $S_{\rm CO} \Delta \nu$ is the flux density expressed in Jy km s$^{-1}$, $D_{\rm L}$ is the luminosity distance in Mpc, and $z$ is the redshift of the galaxy. Here we have assumed the Galactic CO-to-H$_{2}$ conversion factor $X_{\rm CO}$ = $2 \times 10^{20}$ cm$^{-2}$ (K km s$^{-1})^{-1}$ and a constant CO(2-1)/CO(1-0) flux density ratio of 3.2 to convert CO(2-1) flux densities into estimated CO(1-0) values. The measured molecular gas mass values and upper limits are given in Table~\ref{tbl-obsalma}. Our derived cold gas masses are consistent with \citet{Boz} measurements.

Besides five ALMA molecular gas mass measurements we also use previously published results of \citet{Young:11, Su:13, Boz} and \citet{Pulido:17} that have been obtained for galaxies related to the ATLAS${}^{3D}$ survey, individual ETGs, and galaxy clusters, including their BCGs. The ATLAS${}^{3D}$ survey is a volume-limited sample of 260 ETGs that is widely used to study ETG formation and evolution. This sample was observed in CO $J$=1-0 and 2-1 using the IRAM 30 m telescope \citep{Young:11} and 65\% of them were observed in HI using the Westerbork Synthesis Radio Telescope (WSRT) \citep{Morganti:06, Serra:12}. CO emission was detected in 56 targets ($\sim$ 22\%). Here we use 33 CO detections and upper limits of the ATLAS${}^{3D}$ survey. The clusters and their BCGs in \citet{Pulido:17} were observed in CO using the IRAM telescope as well.

\subsection{$M_{\rm mol}-M_{\rm X}$ relation}
We plot molecular gas mass versus X-ray atmospheric gas mass in Fig.~\ref{fig_mm}.  X-ray masses are derived within 10 kpc. This radius was chosen for two reasons. First we wish to obtain a representative mass for the central galaxy that avoids the uncertainties associated with effective radius measurements.  Second, thermodynamic parameters for cluster atmospheres, such as cooling time and entropy, have been reported at this radius to avoid resolution biases when comparing distant to nearby clusters \citep{Rafferty:08}. While this is not an issue for the early-type galaxies studied here, it often is when comparing them to their distant cluster counterparts. In the nearest galaxies we have the opposite issue. Gas mass profiles extend out to only a few kpc before falling off the ACIS camera. In these instances, we have extrapolated their profiles using the linear slope of the last 5 points in log-log space. The BCGs represented as black points in Fig.~\ref{fig_mm} are taken from \citet{Pulido:17}. The remaining points refer to the early-type galaxies analyzed here. The morphological type of each galaxy is indicated: blue circles correspond to ellipticals, magenta points correspond to S0/lenticular galaxies, and the yellow points represent early spirals. Galaxies with recognizable disks are plotted with large shaded diamonds surrounding their native symbol.

Accounting for both detections and upper limits, a trend between molecular gas mass and atmospheric mass is found with correlation coefficient 0.76. The coefficient rises to 0.87 when  the disk-like objects are excluded. The cold gas mass of galaxies with disks, $\sim$ 10$^{7-9}\rm M_{\odot}$, lie an order of magnitude above those without a clear disk, $\sim$ 10$^{6-8}\rm M_{\odot}$. 

Fig.~\ref{fig_mm} clearly shows an increase in scatter about the mean relationship in the lower-mass, early-type galaxies compared to the BCGs.  While much is likely real mass variance, other factors may contribute, the most important being variance in the CO optical depth \citep{Crocker:12}. CO optical depth depends on the gas metallicity and the dynamical state of the gas \citep{Alatalo:13}. Different studies have also adopted different Galactic CO-to-H$_2$ conversion factors to those used in \citet{Pulido:17}, \citet{Boz}, and \citet{Young:11} depending on perceived conditions in the gas. However, these differences are generally a factor of a few or so, while the molecular gas mass at a given X-ray atmospheric gas mass in Fig.~\ref{fig_mm} spans more than two orders of magnitude. Therefore most of the scatter in the relationship is intrinsic.

A linear fit to the $M_{\rm mol}-M_{\rm X}$ relation was performed using only detections of cold gas masses.  A bivariate correlated error and intrinsic scatter (BCES) routine \citep{Akritas:96} was used to perform linear modeling in log space. Parameter errors are determined using 10,000 iterations of Monte Carlo bootstrap re-sampling. Using the BCES routine we found $M_{\rm mol} \propto M_{\rm X}^{1.4\pm0.1}$. 

The best-fit is shown in Fig.~\ref{fig_mm} as a black solid line. Although a clear dependence of $M_{\rm mol}$ on $M_{\rm X}$ if found for clusters (BCGs), little correlation is found for the early-type galaxies alone. Furthermore, the scatter about the mean of the early-type galaxies exceeds that of the BCGs by a factor of 2. The relation between cold gas mass and X-ray mass for clusters alone follows the steeper form, $M_{\rm mol} \propto M_{\rm X}^{1.6\pm0.1}$. To explore this relationship further, we include the molecular mass upper limits using survival analysis.

\subsubsection{Survival analysis}
Survival analysis is a powerful tool to estimate the likelihood and form of a relationship by evaluating both detections and upper limits \citep{statsu, statsu1, statsu2}  
The likelihood is expressed as
\begin{equation}
 L = \prod_{i=1}^n Prob[t_i,\delta_i] = \prod_{i=1}^{n} [f(t_i)^{\delta_i}][1 - S(t_i)]^{1-\delta_i},
\end{equation}
where, $t_i = min(x_i,c_i)$, $x_i$ are the detected values, $c_i$ are the upper limits, $\delta_i$ is 0 for upper limits and 1 for detections. The $f_i$ and $S_i$ are the likelihoods for detections and upper limits, respectively. In the case of Normal (Gauss) distribution, the likelihood for the detected values can be expressed as
\begin{equation}
 f(x) = \frac{1}{\sqrt{2\pi}\sigma} \exp{\left[ -\frac{1}{2} \left( \frac{x-\mu}{\sigma}\right)^2 \right]},
\end{equation}
where, $x$ and $\sigma$ are the detected values of molecular gas mass and their errors, while $\mu$ is an assumed model of fitting. The likelihood for upper limits, $S(x)$, can be expressed as 
\begin{equation}
 S(x) = 1-\frac{1}{2} \left[ 1 + {\rm erf} \left( \frac{x-\mu}{\sqrt{2}\sigma} \right) \right],
 \label{eqsx}
\end{equation}
where $\rm erf$ is the error function, $\rm erf(x) = \frac{2}{\sqrt{\pi}} \int_0^x{e^{-t^2}}dt$. In the case of Eq.~\ref{eqsx} we use $x = ({\rm upper\ limit})_i$ and $\sigma = ({\rm upper\ limit})_i/3$. We assume $\sigma$ as ({\rm upper limit})$_i$/3 since our upper limits were defined for 3$\sigma$ confidence level. 

\begin{figure}    
\includegraphics[width=0.49\textwidth]{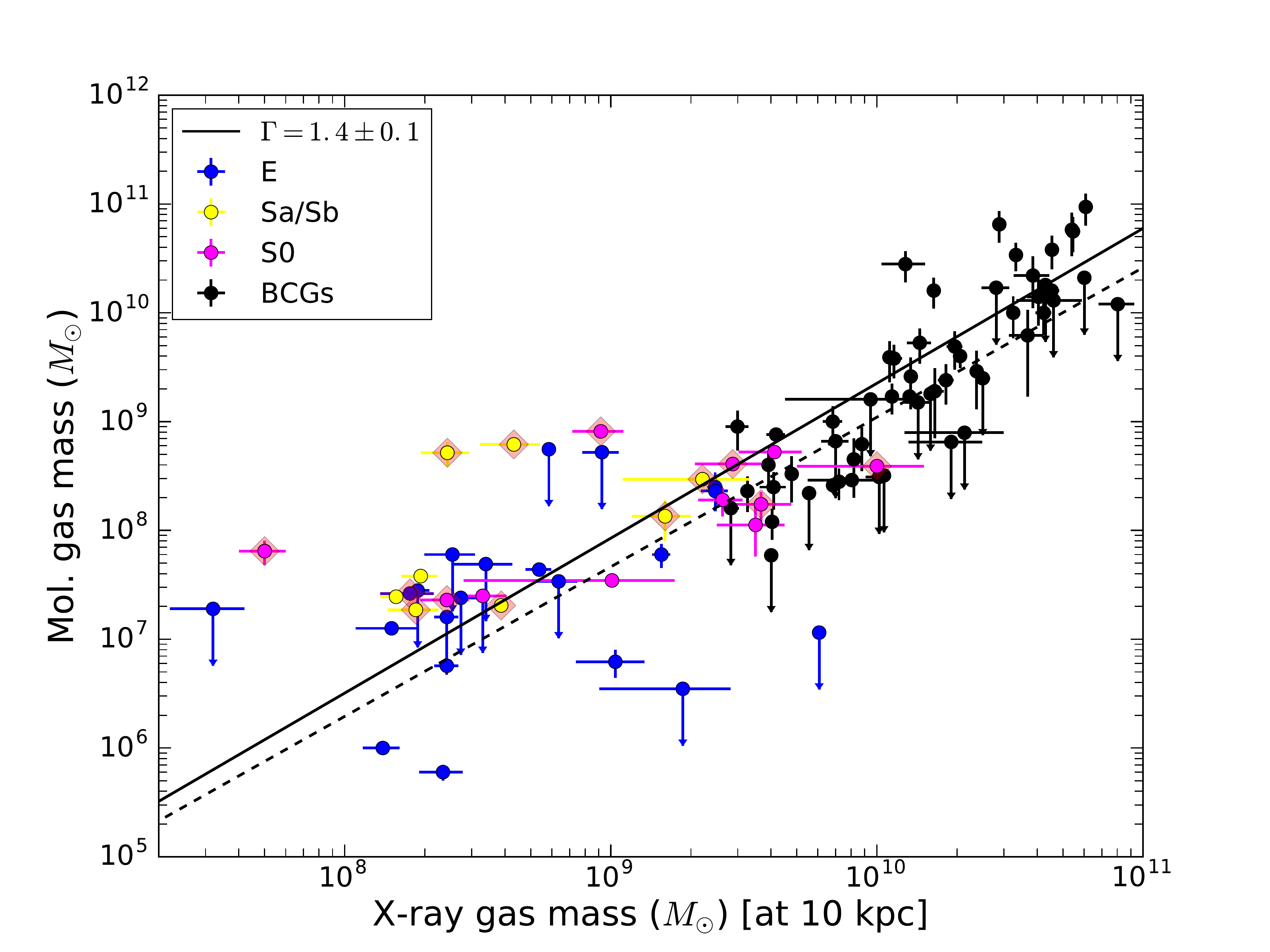}
\caption{The relation of cold molecular gas mass to X-ray gas mass. The morphological type of each galaxy is indicated: blue circles correspond to ellipticals, magenta points correspond to S0/lenticular galaxies, and the yellow points represent early spirals. Galaxies with recognizable disks are plotted with large shaded diamonds surrounding their native symbol.}
\label{fig_mm}
\end{figure}

\begin{figure}[t!]
\centering
\includegraphics[width=0.46\textwidth]{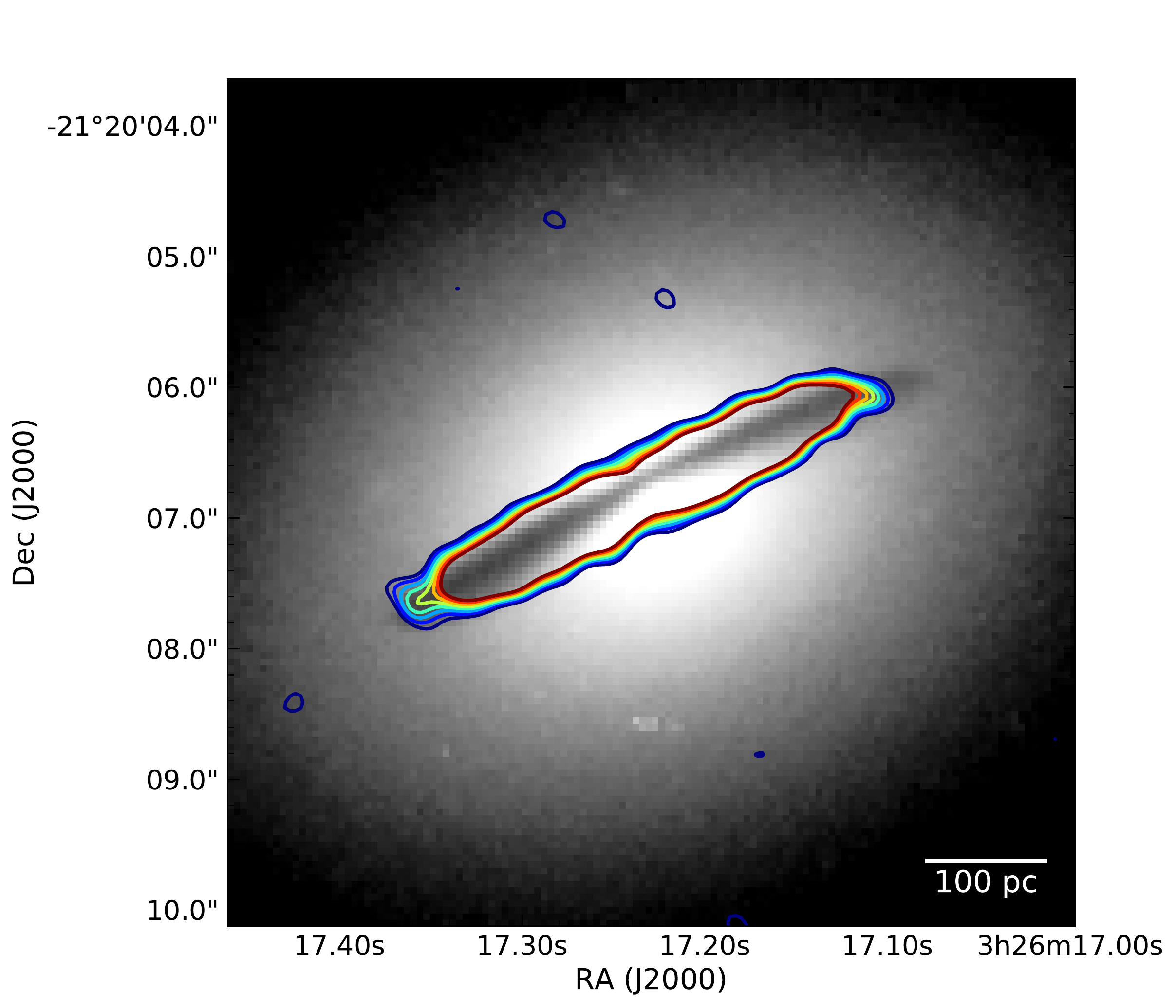}\\
\includegraphics[width=0.49\textwidth]{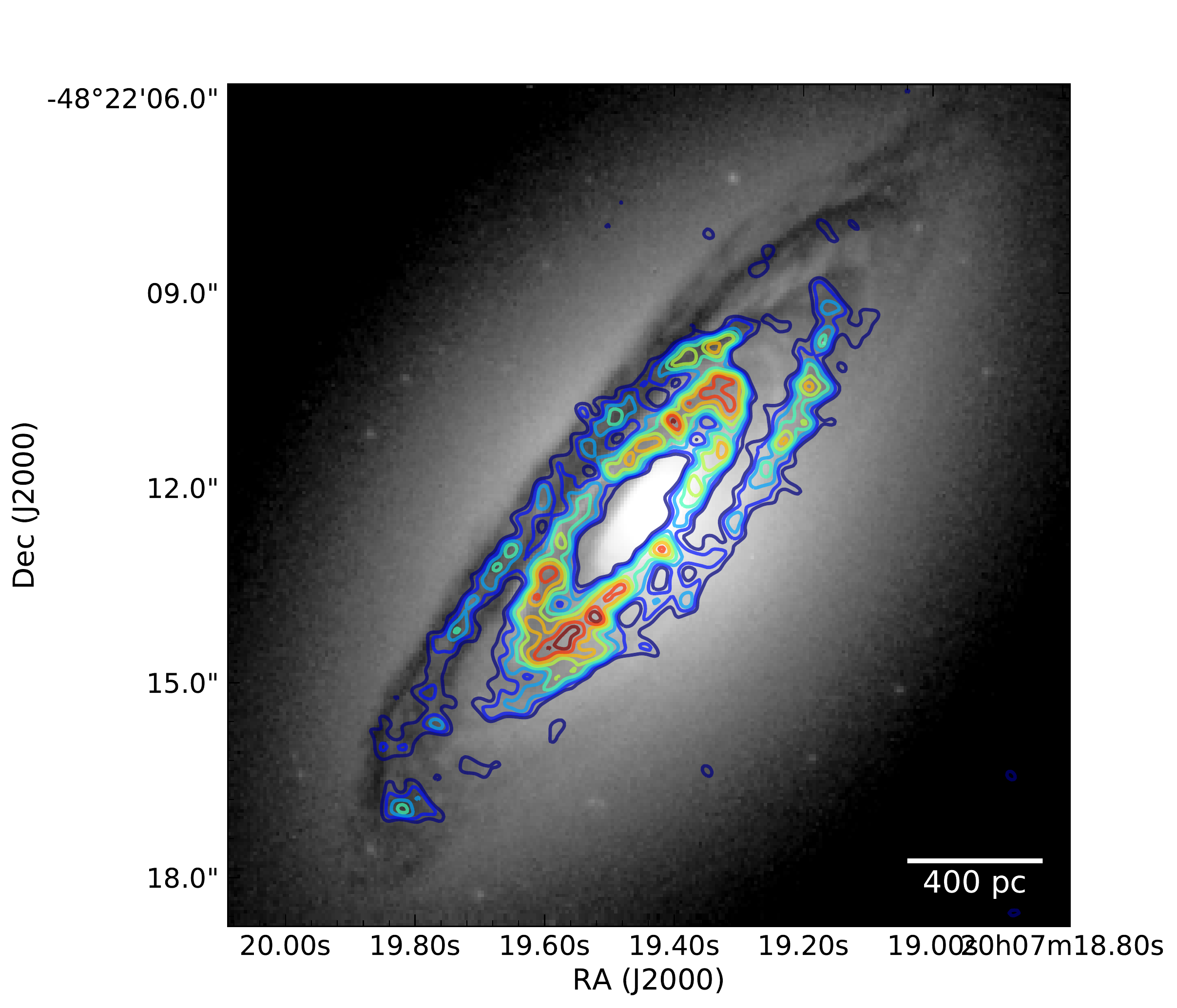}
\caption{\textit{HST} (F814W) linear gray scale images of NGC 1332 (\textit{top}) and NGC6861 (\textit{bottom}) overlaid with ALMA CO(2-1) emission line contours from integrated intensity maps. In both images, contours start at 3$\sigma$ level and increase linearly.}
\label{fig_alma}
\end{figure}

For the model, $\mu$, we use the power law of the form of $A\times\left(\frac{x}{x_0} \right)^{\alpha}$. We find slightly shallower relation, $M_{\rm mol} \propto M_{\rm X}^{1.37\pm0.21}$, than those obtained with BCES routine that included only detections. The fit is presented in Fig.~\ref{fig_mm} as a dashed line. Galaxies with gas disks have the largest residuals. Two disk systems are shown in Fig.~\ref{fig_alma}. Avoiding these points, the $M_{\rm mol}-M_{\rm X}$ relation follows a steeper relation with less scatter, $M_{\rm mol} \propto M_{\rm X}^{1.48\pm0.15}$. 

Fits were calculated separately for the early-type galaxies and cluster central galaxies. Including cluster central galaxies alone we find  $M_{\rm mol} \propto M_{\rm X}^{1.51\pm0.13}$.  For early-type galaxies we find $M_{\rm mol} \propto M_{\rm X}^{0.48\pm0.81}$.  The early-type galaxies alone reveal no correlation, implying two possibilities:  Either the molecular gas in early-type galaxies is unrelated to atmospheric gas, or the range of molecular gas mass and atmospheric mass in the early-type galaxy sample is too limited relative to the variance to reveal the true underlying relationship.  Statistics alone will not discriminate between these possibilities.  
However, other factors argue in favour of a physical trend from centrals in clusters to early type galaxies.  

First, molecular gas rich galaxies all have similarly short atmospheric cooling times lying below 1 Gyr at 10 kpc (see below), while gas poor systems have atmospheric cooling times exceeding 1 Gyr at 10 kpc. This is true in both early-type galaxies and in cluster centrals where molecular gas masses are too high to have arrived in mergers. 

\begin{figure}    
\centering
\includegraphics[width=0.49\textwidth]{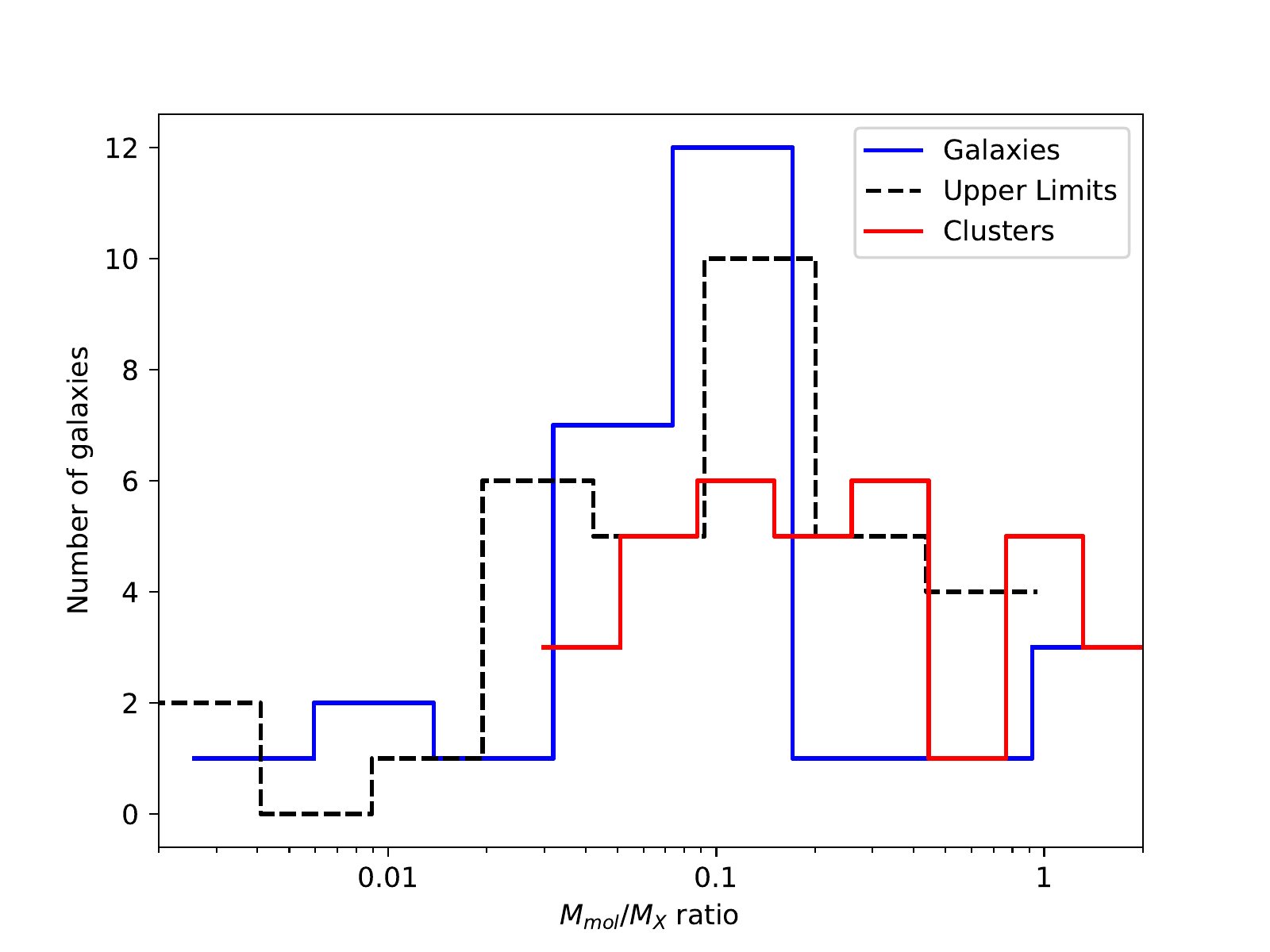}
\caption{The fraction of molecular gas mass in X-ray gas mass for both high- and low-mass systems. The X-ray gas mass is taken at 10 kpc.}
\label{fig_hist}
\end{figure}

Second, the ratio between atmospheric gas mass and molecular gas mass of the two populations is similar.   Fig.~\ref{fig_hist} shows the mass ratios of molecular to atmospheric gas for both cluster centrals and early-type galaxies within a 10 kpc radius, where most molecular gas lies.   This ratio lies between $10-20\%$ on average but with large variance. The distribution of upper limits is broadly consistent with the detections, indicating they are likely within factors of several of their true values.  


This level of concordance would be difficult to understand if the origins of molecular clouds in BCGs and early-type galaxies were dramatically different.  For example if the cold gas arrived exclusively by mergers or filaments, there is no obvious reason the atmospheric mass would exceed the molecular gas masses by such a large factor.  Nevertheless, this figure indicates that additional CO measurements are needed to convert the upper limits into detections before drawing firm conclusions about the relationships between cold gas and atmospheric cooling in early-type galaxies.

\subsection{Density, Temperature, and Luminosity}
\begin{figure}    
\centering
\includegraphics[width=0.47\textwidth]{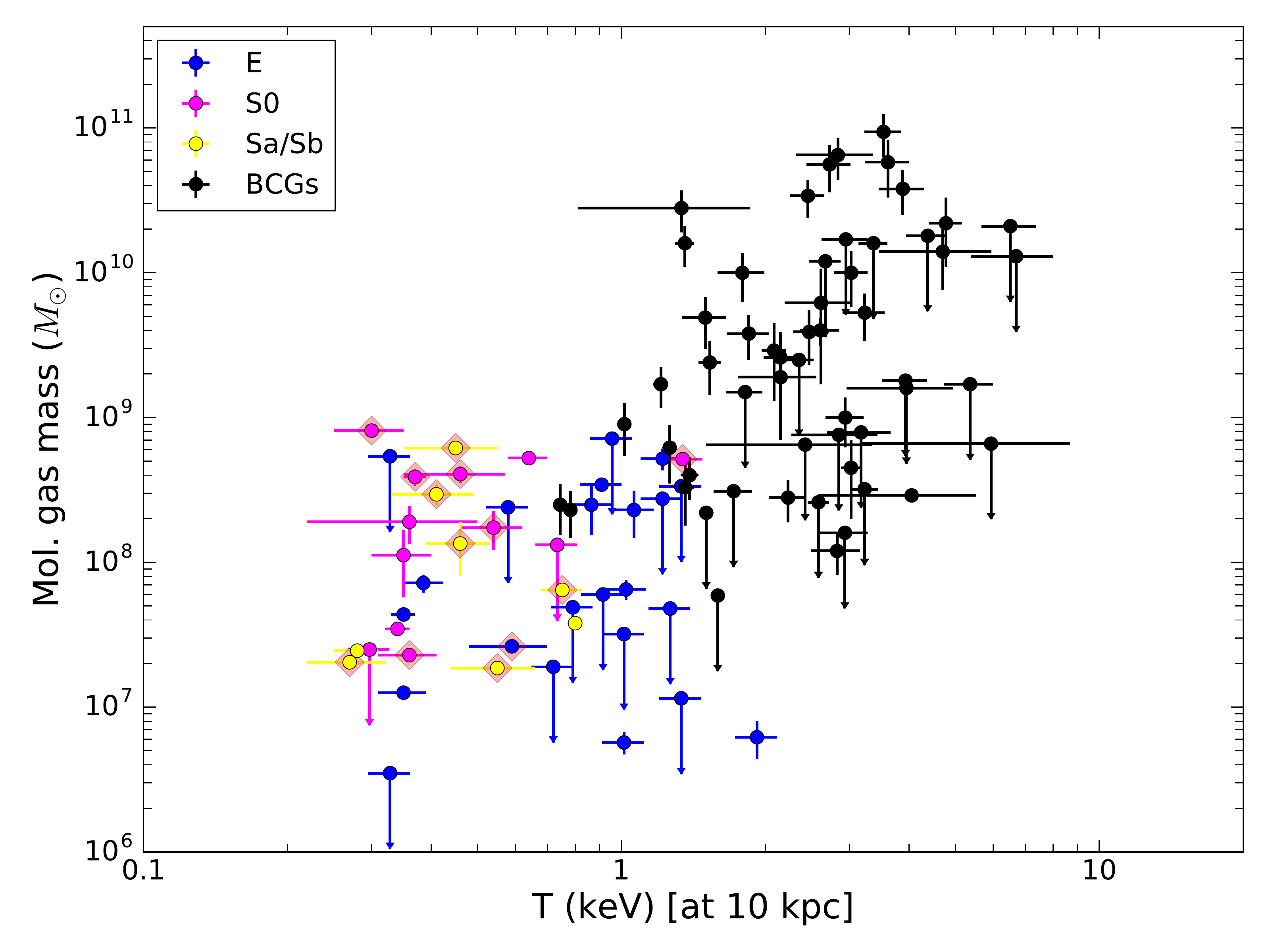}
\includegraphics[width=0.47\textwidth]{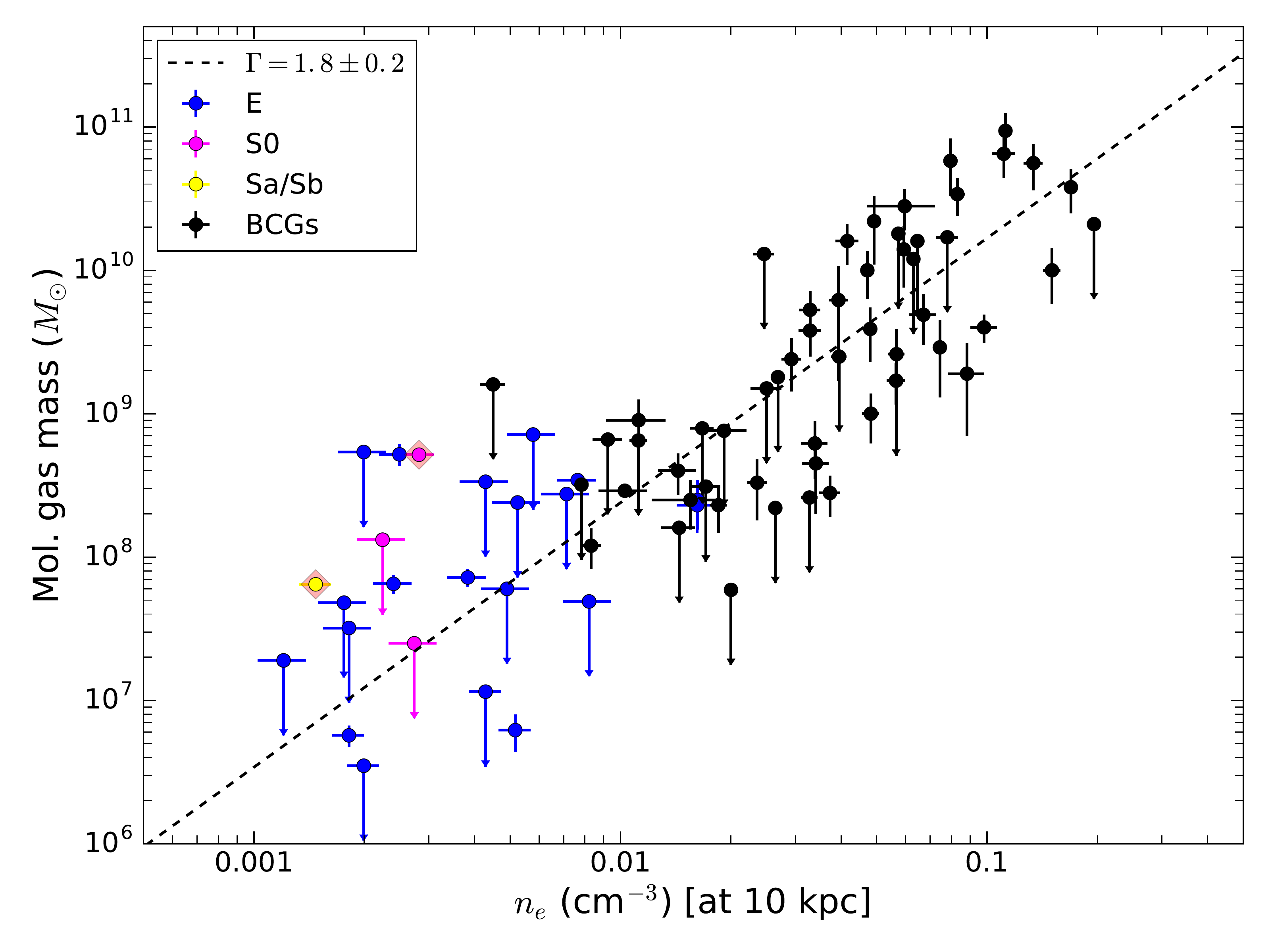}
\caption{The molecular gas mass versus temperature (upper panel) and electron density at 10 kpc (lower panel). The $M_{\rm mol}-T$ relation includes the spectral temperature measurements presented in Table~\ref{tab3}.}
\label{fig_mol_params}
\end{figure}

The dependence of molecular gas mass on atmospheric temperature, electron density, entropy, luminosity and the minimum value of the \tctff\ ratio are investigated here. The upper panel of Fig.~\ref{fig_mol_params} shows molecular gas mass versus mean atmospheric gas temperature.  The trend indicates that higher temperature systems generally host more massive reservoirs of cold gas. This trend is not surprising as it indicates that hotter, hence more massive systems, attract and retain more massive molecular gas reservoirs. The scatter in molecular gas mass at a given temperature spans is roughly two decades, indicating mass alone does not determine the level of the molecular gas reservoir.  
 
An indication that atmospheric gas density is a significant factor is shown in the lower panel of Fig.~\ref{fig_mol_params}. The figure shows a trend between molecular gas mass and atmospheric gas density measured at an altitude of 10 kpc. This trend was found by \citet{Pulido:17} for cluster central galaxies. Their data in plotted to the upper right in Fig.~\ref{fig_mol_params}. The trend extends to early-type galaxies, plotted to the lower left, albeit with greater scatter than the cluster central galaxies. It is not clear whether the greater scatter is intrinsic or due to the large number of upper limits.

A linear fit to $M_{\rm mol}-n_{\rm e}$ relation was performed using survival statistics. We find that this relation follows a power law scaling as $M_{\rm mol} \propto n_{\rm e}^{1.8\pm0.2}$. The scatter of $M_{\rm mol}-n_{\rm e}$ relation is 0.34 dex while the correlation coefficient is 0.86. The fit is shown as a dashed line in Fig.~\ref{fig_mol_params}. Separate analyses were performed for the clusters and early-type galaxies alone. The cluster central (black points) relation follows a power-law scaling as $M_{\rm mol} \propto n_{\rm e}^{2.2\pm0.4}$. This is slightly steeper compared to the full $M_{\rm mol}-n_{\rm e}$ relation. However, both are consistent within their uncertainties. In the case of low-mass systems (blue, magenta, and yellow points), we find $M_{\rm mol} \propto n_{\rm e}^{1.5\pm1.1}$, which is shallower than both the entire sample and cluster central galaxies alone. However, the slope of $M_{\rm mol}-n_{\rm e}$ relation for early-type galaxies is consistent, within uncertainties, with the full sample. The scatter of $M_{\rm mol}-n_{\rm e}$ relation for BCGs is 0.24 dex, while for low-mass systems is 0.56 dex.

\begin{figure}    
\centering
\includegraphics[width=0.52\textwidth]{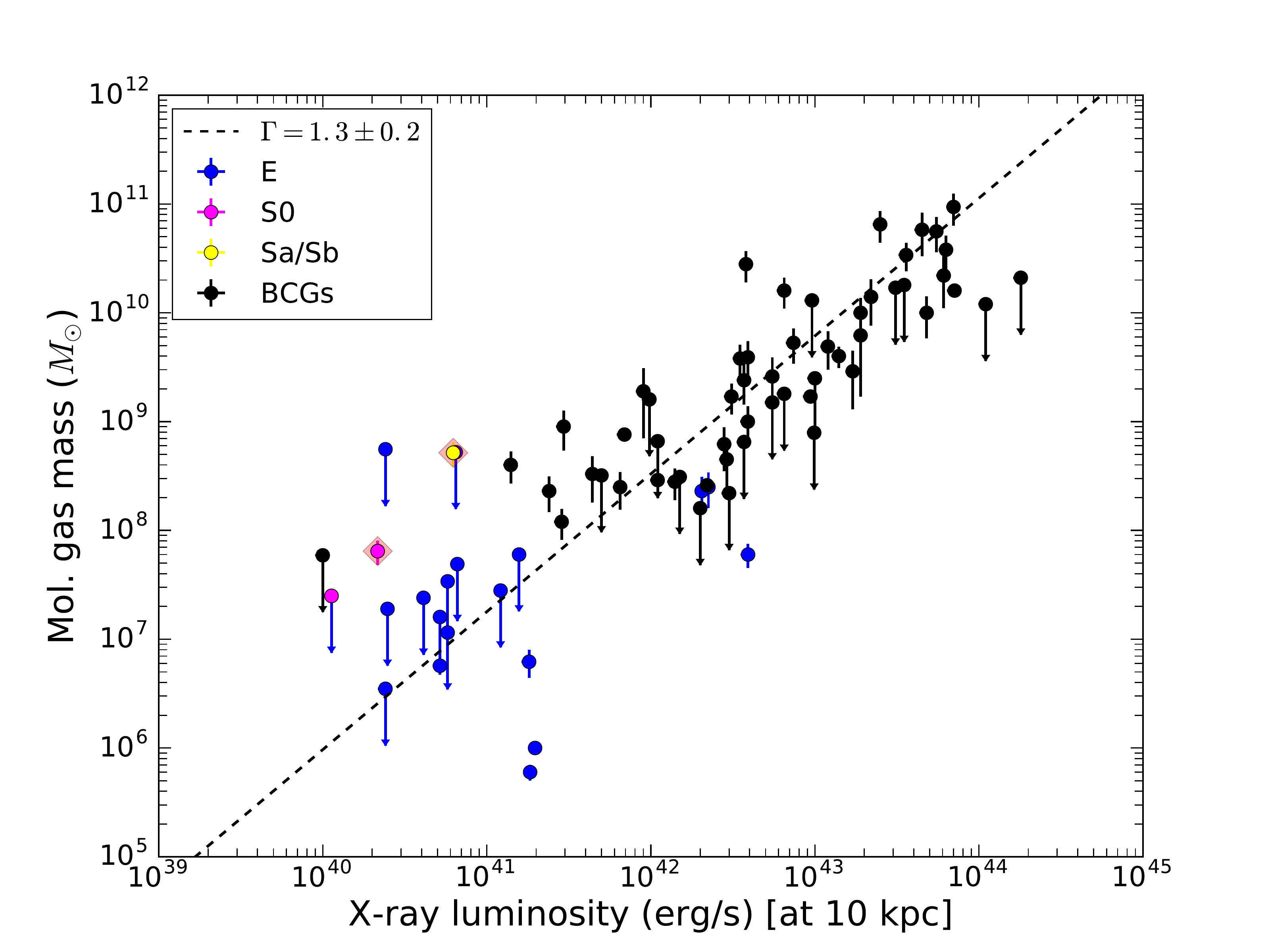}
\caption{The molecular gas mass versus X-ray luminosity defined at 10 kpc. The $M_{\rm mol}-L_{\rm X}$ relation is fitted by a power-law model using survival analysis.}
\label{fig_mol_lx}
\end{figure}

The $M_{\rm mol}-L_{\rm X}$ relation for the entire sample was constructed. The X-ray luminosity within 10 kpc was computed as, $L_{\rm X} = 4\pi D_{\rm L}^{2}f_{\rm X}$, where $f_{\rm X}$ is the X-ray flux obtained from spectral fitting. The $M_{\rm mol}-L_{\rm X}$ relation is shown in Fig.~\ref{fig_mol_lx}.  The survival fit for the entire sample yields  $M_{\rm mol} \propto L_{\rm X}^{1.3\pm0.2}$. The scatter is only 0.25 dex, demonstrating a tight dependence. The separate fits for the early-type galaxies and cluster central galaxies alone yields $M_{\rm mol} \propto L_{\rm X}^{0.5\pm0.8}$ and $M_{\rm mol} \propto L_{\rm X}^{1.2\pm0.1}$, respectively.  Therefore, the trends are consistent with each other, albeit with large scatter in the early-type galaxies. The $M_{\rm mol}-L_{\rm X}$ relation for early types alone is consistent with \citet{OSullivan:18}, who likewise found no clear correlation between CO emission and the absence or presence of a hot intra-group medium.

\begin{figure}    
\includegraphics[width=0.5\textwidth]{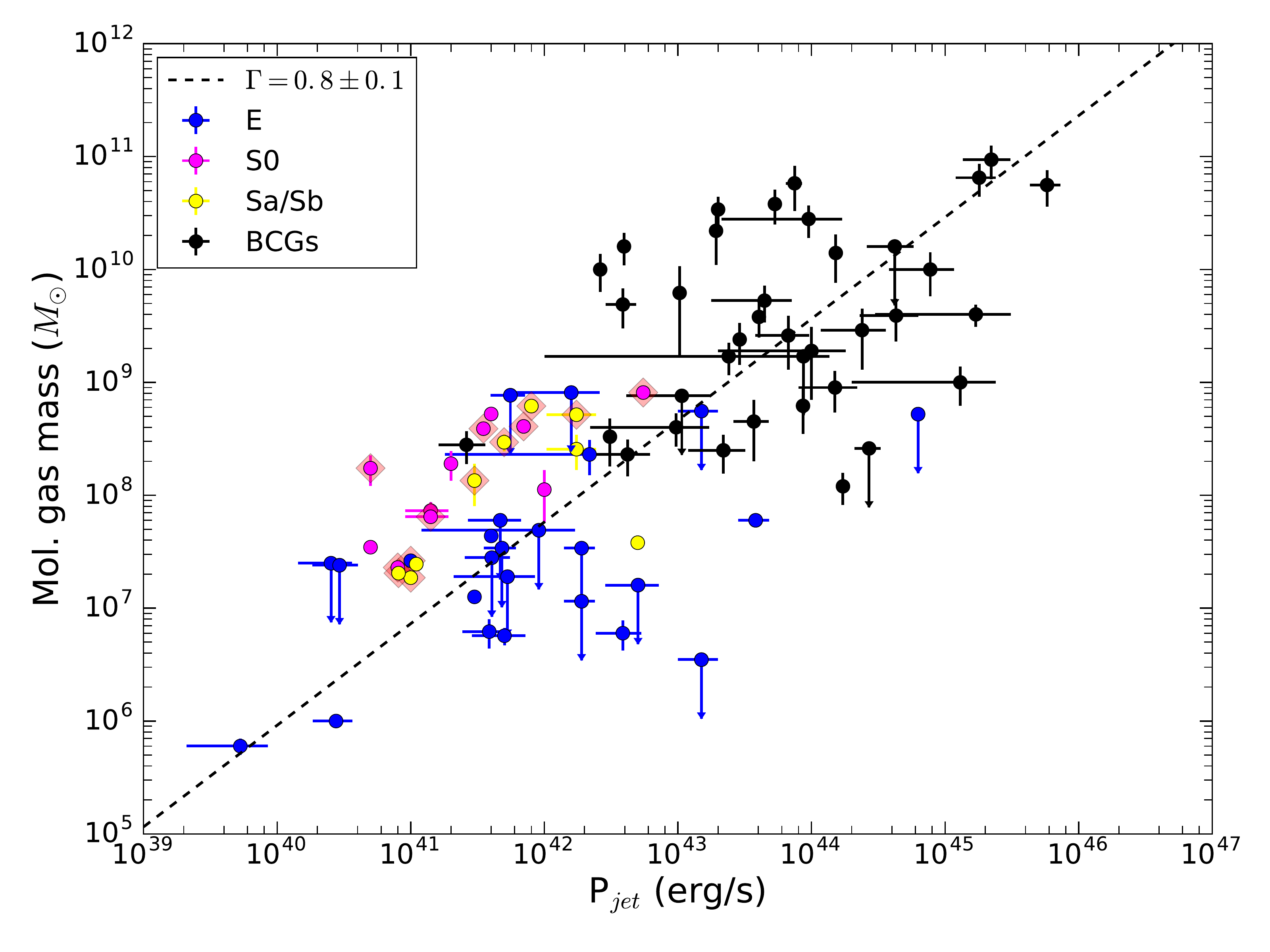}
\caption{The relation of cold molecular gas mass versus AGN jet power. The dashed line corresponds the survival fitting.}
\label{fig_mmcav}
\end{figure}


\subsection{Discussion}
\subsubsection{Does Molecular Gas fuel Radio/AGN Feedback?}
Maintaining balance between heating and cooling of hot atmospheres across a large range of halo mass requires a reliable fuel supply to the AGN. Bondi accretion of the hot atmosphere onto the central black hole would be a feasible fuel supply in early-type galaxies \citep{Rafferty:06, Allen:06, Narayan:11, Hardcastle:06}. However, Bondi accretion would be unable to fuel the most powerful AGN \citep{Russell:13a} in cluster central galaxies \citep{Hardcastle:06, Rafferty:06, McNamara:11}. The CO detections and upper limits in cluster central galaxies and early-type galaxies are easily sufficient to fuel radio AGN.  But no trend between molecular gas and radio AGN power has been found. 
 
Using a recent X-ray cavity analysis, we adopt the scaling relation between radio power and AGN jet power \citep{Cavagnolo:10} to explore this trend. Radio power was calculated using the relation
\begin{equation}
P_{\nu_0} = 4\pi D_{\rm L}^2(1+z)^{\alpha-1} S_{\nu_0} \nu_0,
\end{equation}
where $S_{\nu_0}$ is the flux density at the observed frequency, $\nu_0$, $z$ is the redshift, $D_{\rm L}$ is the luminosity distance, and $\alpha$ is the radio spectral index. The radio flux densities, $S_{\nu_0}$, were taken from the NRAO (National Radio Astronomy Observatory) VLA Sky Survey (NVSS) \citep{Condon:98}. We found no radio fluxes for 22 galaxy clusters. We assume spectral index $\alpha$ = 0.8 and ${\nu_0}$ = 1.4 GHz. We estimate the AGN mechanical power in BCGs to be $10^{43-45}~\rm erg~s^{-1}$, which is two to three orders of magnitude higher than those in the early-type galaxy sample. 

\begin{figure*}
\includegraphics[width=0.335\textwidth]{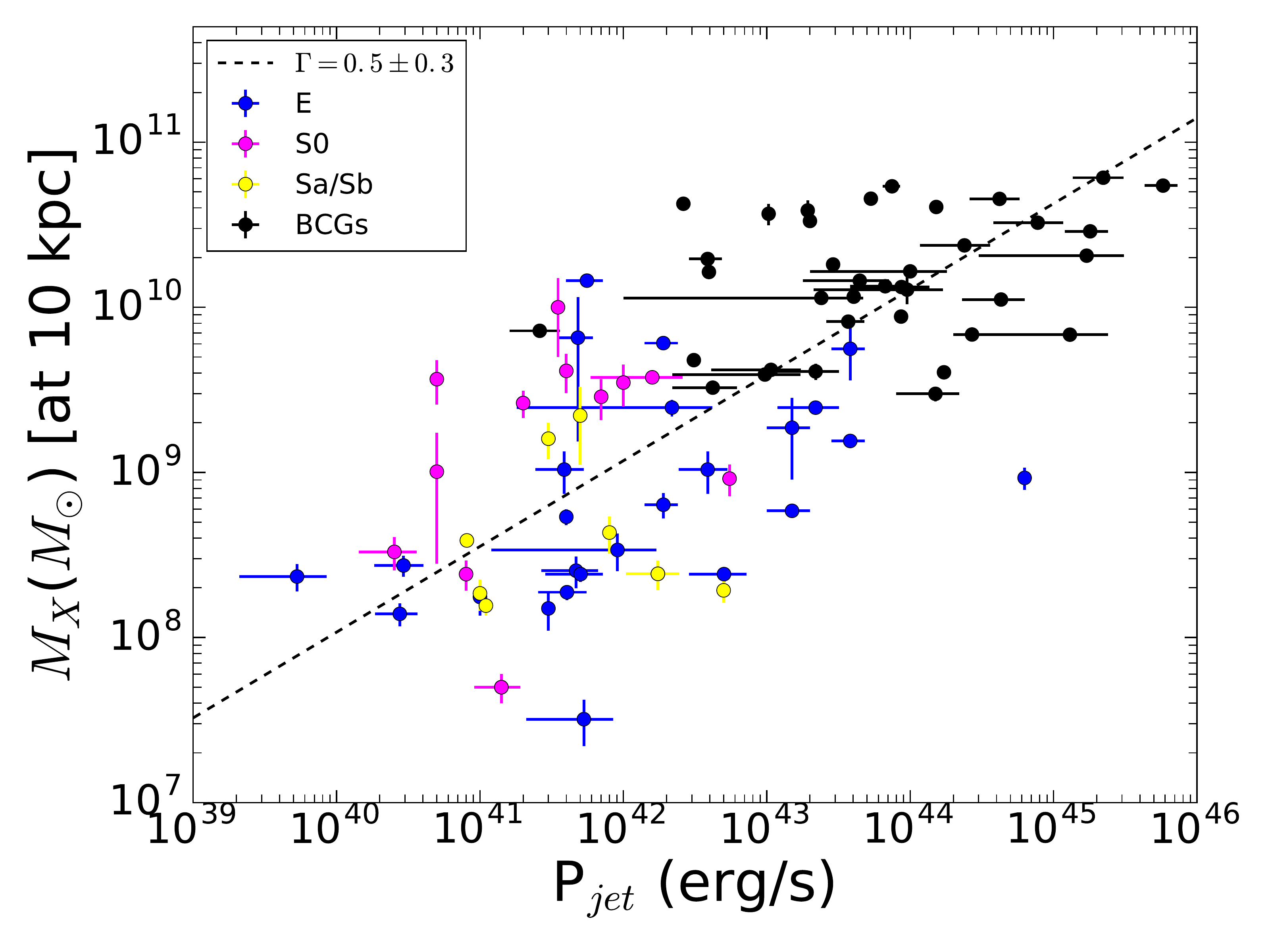}
\includegraphics[width=0.335\textwidth]{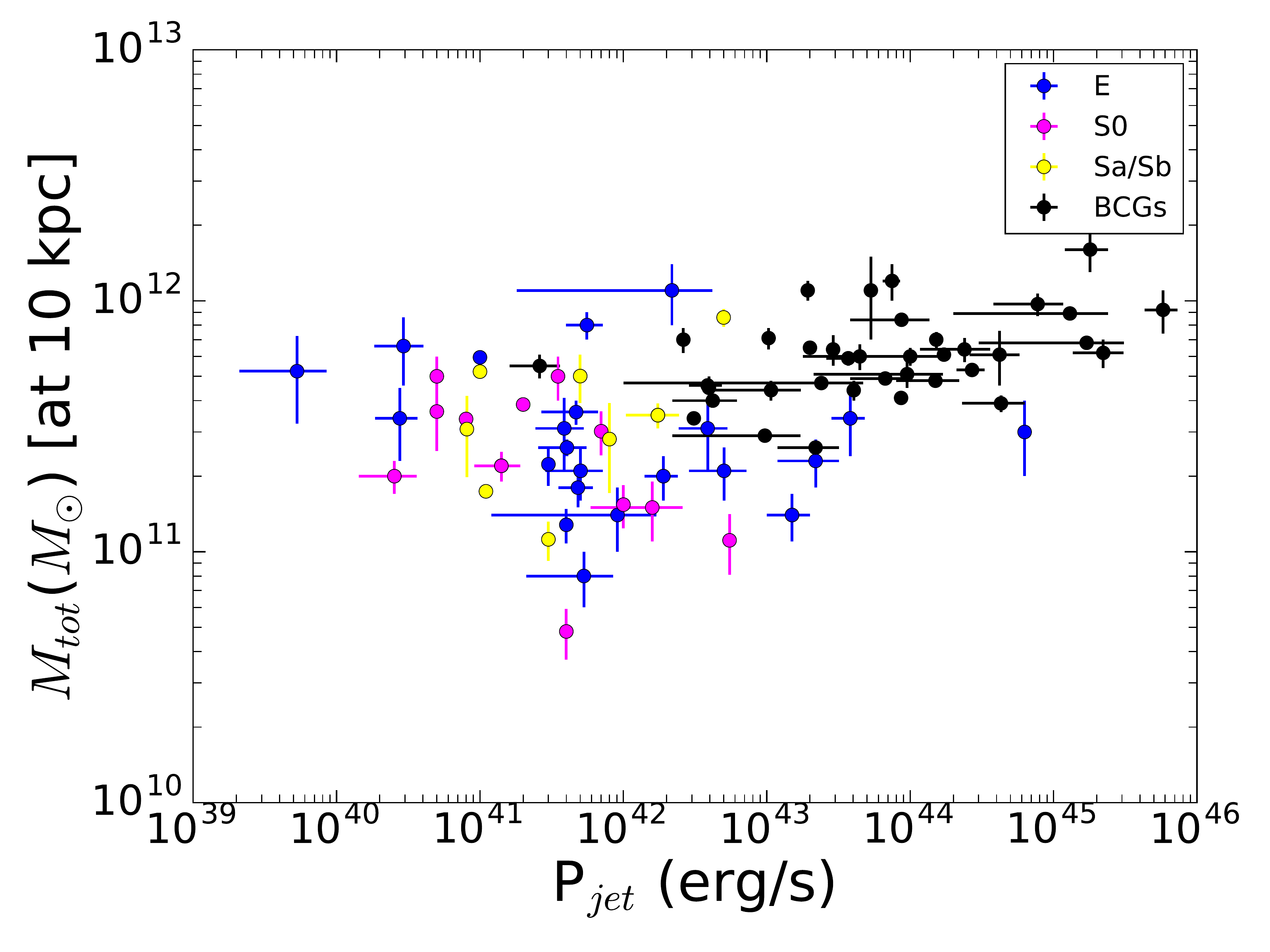}
\includegraphics[width=0.335\textwidth]{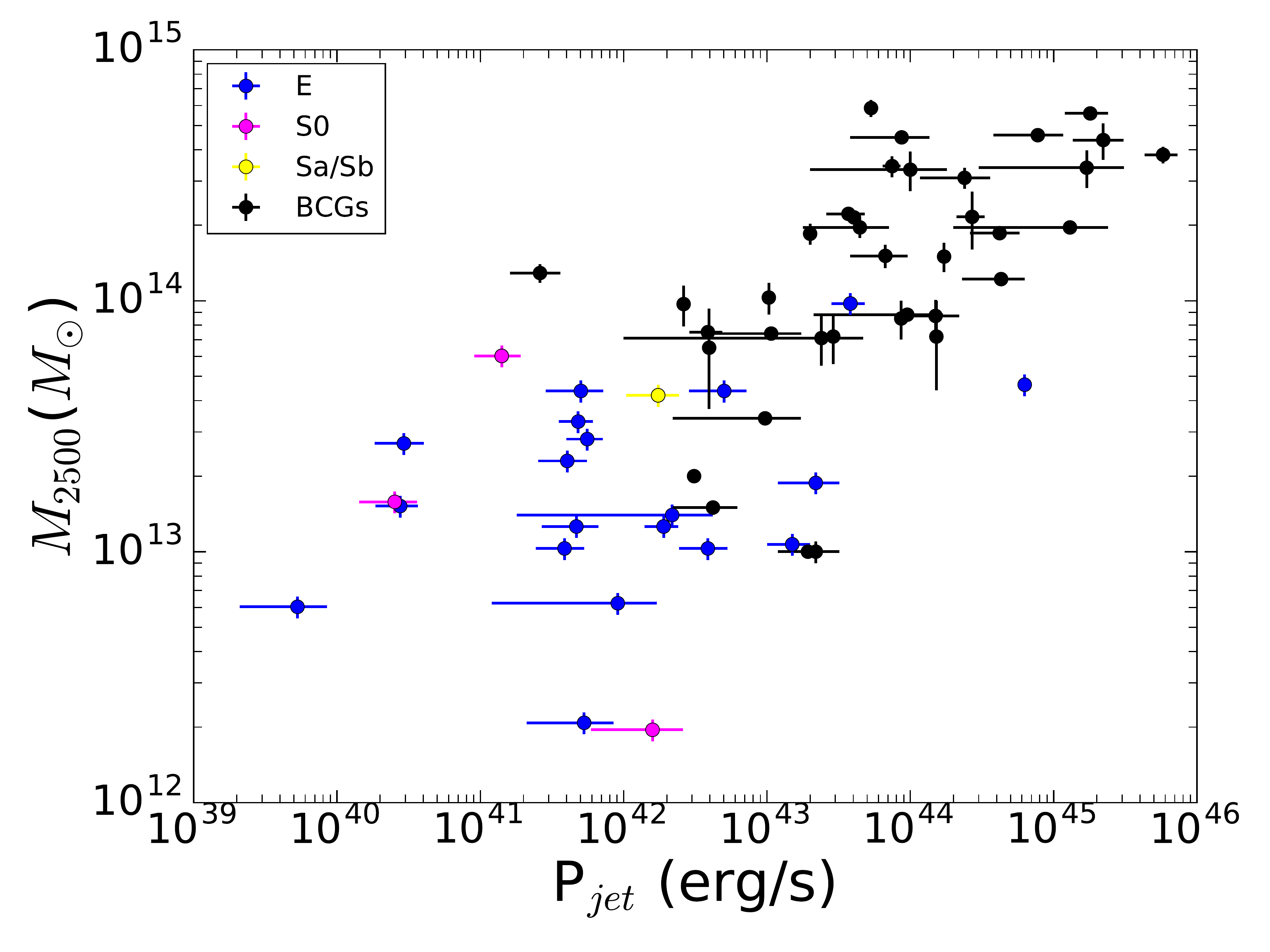}
\caption{The X-ray gas mass, total mass, and $M_{2500}$ versus the radio AGN jet power (from left to right).}
\label{fig_cavmass}
\end{figure*}

The correlation between molecular gas mass and radio jet power is presented in Fig.~\ref{fig_mmcav}.  A weak trend over five decades in molecular gas mass and nearly six decades in jet power is found.  The scaling is nearly linear, following \mm $\propto P_{\rm jet}^{0.8\pm0.1}$ with a standard deviation of 1.21 dex. However, the correlation coefficient 0.64 indicates the trend is at best marginally significant.  For a given molecular gas mass, one finds roughly three decades of variation in jet power, and conversely so. A similar variance was noted by \citet{McNamara:11} in cluster central galaxies alone. 

To further evaluate the degree of correlation between these quantities we explore the influence of $D^2$ factor on both axes. We generated random redshift, distance, and flux datasets with the correct spread given by error bars over the observed range. We recovered the scatter in the observed $M_{mol}$-$P_{jet}$ relation but the simulated slope, 1.0$\pm$0.03,
was steeper than the observed slope $0.8\pm0.1$, at about the $2\sigma$ significance level.  This result is consistent with the correlation coefficient which indicated only marginal significance.

One would naively expect a large scatter, hence a weak correlation between jet power and the level of the molecular gas reservoir.  Only a tiny fraction of the molecular gas reservoirs shown in Fig.~\ref{fig_mmcav} would be required to fuel the AGN at the levels observed in this figure \citep{McNamara:11}. Furthermore, the molecular gas is distributed on much larger scales than the central black hole's sphere of influence. So the timescale for the most of the molecular gas to accrete onto the central black hole dramatically exceeds the age of the AGN. Only the mass on small scales is currently participating in fueling the AGN. Therefore, the trend indicates that systems that on average have higher AGN jet power also contain larger molecular gas reservoirs available to fuel the AGN.  

Correlations between jet power, atmospheric gas mass within 10 kpc, total mass within 10 kpc, and total mass within $R_{2500}$ are explored in Fig.~\ref{fig_cavmass}. Weak correlations are found between AGN jet power and both atmospheric gas mass (correlation coefficient 0.48) and total mass, $M_{2500}$ (correlation coefficient 0.44).  $M_{2500}$ for the early-type galaxies was determined using the $M_{2500}-T$ scaling relation of \citet{Vikhlinin:06}. This approach was adopted because we are unable to trace the gas density and temperature out to this distance for the nearest galaxies. In contrast, cluster central masses, $M_{2500}$, can be measured directly \citep{Hogan:17a, Pulido:17}. The extrapolation to $R_{2500}$ contributes scatter to the early-type galaxy measurements. However, a correlation between molecular gas mass and total mass within $R_{2500}$ is apparent. In contrast, no correlation is found between AGN jet power and total mass within 10 kpc. The absence of correlation is likely due to the dominant stellar component in the total mass within 10 kpc, which would not participate in fueling the AGN. It is interesting, nevertheless, that $M_{2500}$ and the atmospheric gas are correlated with AGN jet power.

These trends are consistent with the large-scale mass dependence of AGN power in systems with central cooling times shorter than 1 Gyr \citep{Main:17} in clusters. This new result extends Main's to early-type galaxies. This correlation indicates that AGN power, which emerges from processes near to the event horizon are regulated by conditions on vastly larger scales.

\subsubsection{Cold Gas in Early-type Galaxies}
While the origin of molecular gas in BCGs is almost certainly cooling from the hot atmosphere, its origins in early-type galaxies is less clear. Internal origins include stellar ejecta and atmospheric cooling, or externally-accreted gas from mergers of gas-rich galaxies. We discuss these ideas in turn.

\subsection{Atmospheric Cooling}

Using the {\it Herschel} Observatory, \citet{Werner:14} showed that the molecular cooling lines of [CII] and [OI] correlate with nebular emission in some early-type galaxies. The systems with nebular emission demonstrate flatter entropy profiles than those without nebular emission, tentatively connecting the cold gas to cooling from the hot atmospheres. They also found that thermally unstable systems indicated by the ``Field Criterion'' are likely to have [CII] emission from warm molecular gas. This again is consistent with cooling from the hot atmosphere. Similar results have been found in cluster BCG atmospheres \citep{Voit:08, Rafferty:08, Cavagnolo:08}. 

The \citet{Werner:14} study was extended to 49 nearby elliptical galaxies by \citet{Werner:18new}, who found a similar connection between nebular emission and cooling hot atmospheres, indicating a common origin between the hot atmosphere and molecular gas. Lakhchaura's sample significantly overlaps our own. The correlations we find here between the thermodynamic properties of hot atmospheres and CO mass, in combination with the Werner and Lakhchaura results,  point to cooling from hot atmospheres as a significant, perhaps the most significant, source of molecular gas in early-type galaxies.

\begin{figure*}    
\centering
\includegraphics[width=0.33\textwidth]{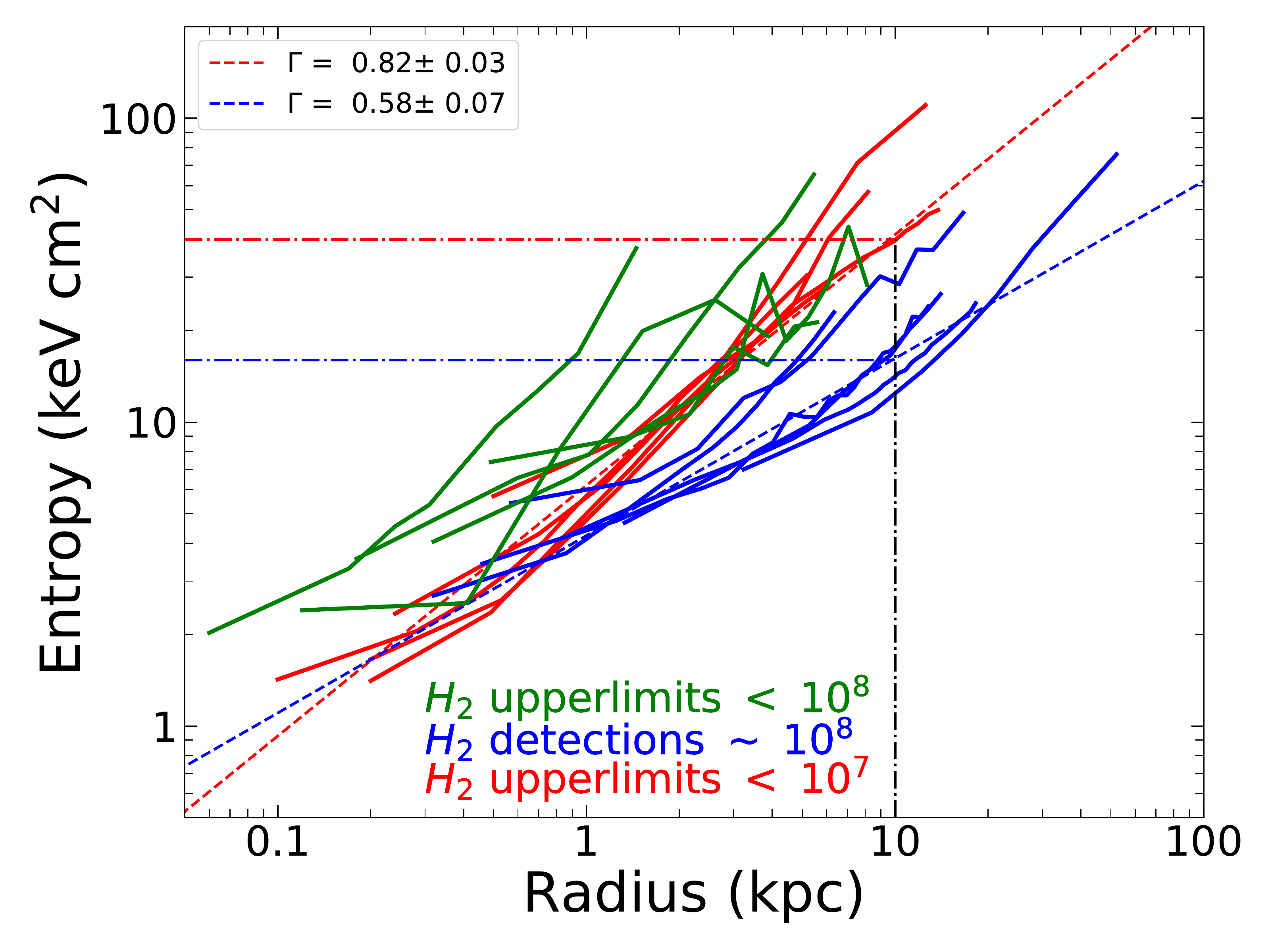}
\includegraphics[width=0.33\textwidth]{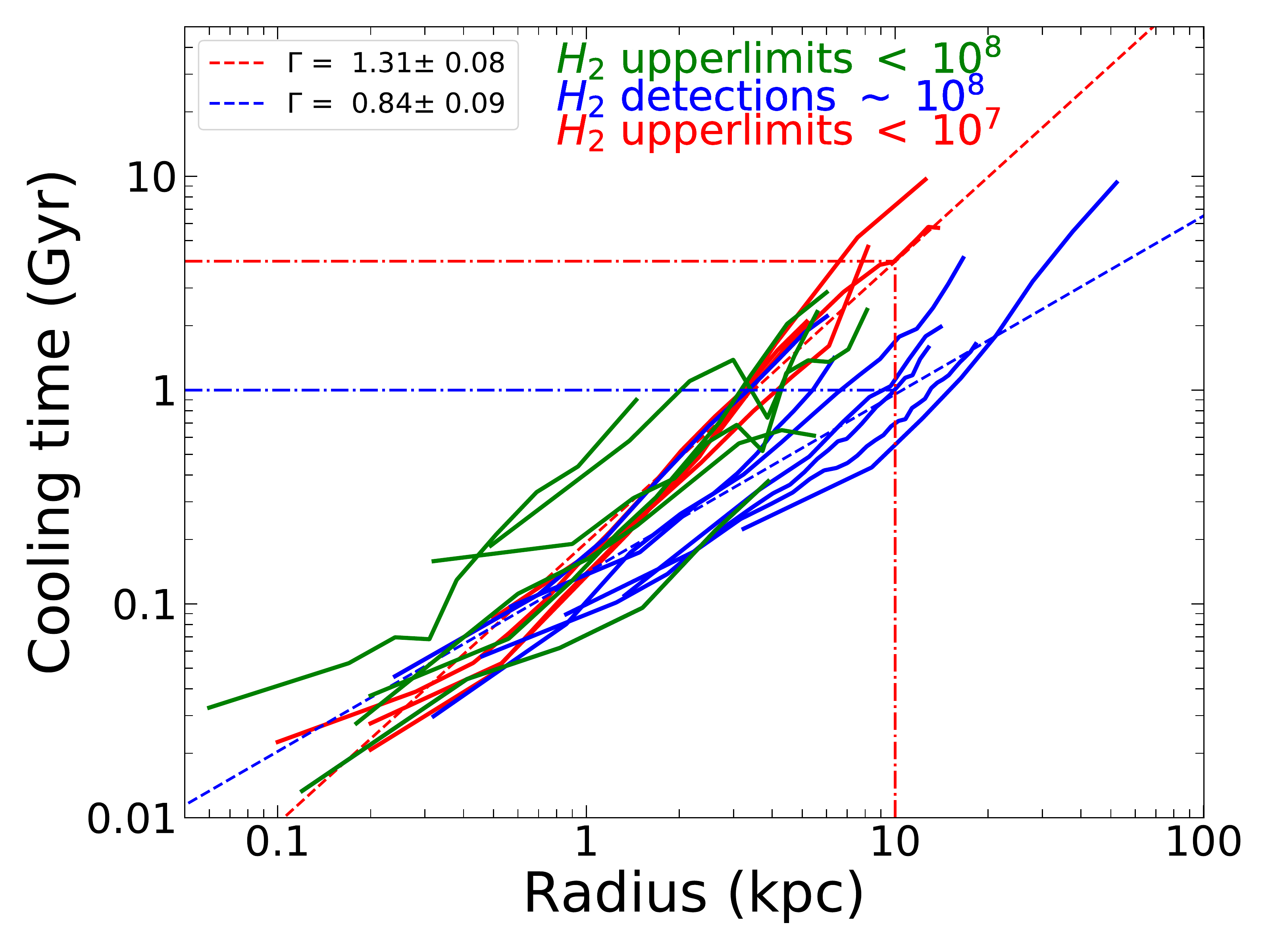}
\includegraphics[width=0.33\textwidth]{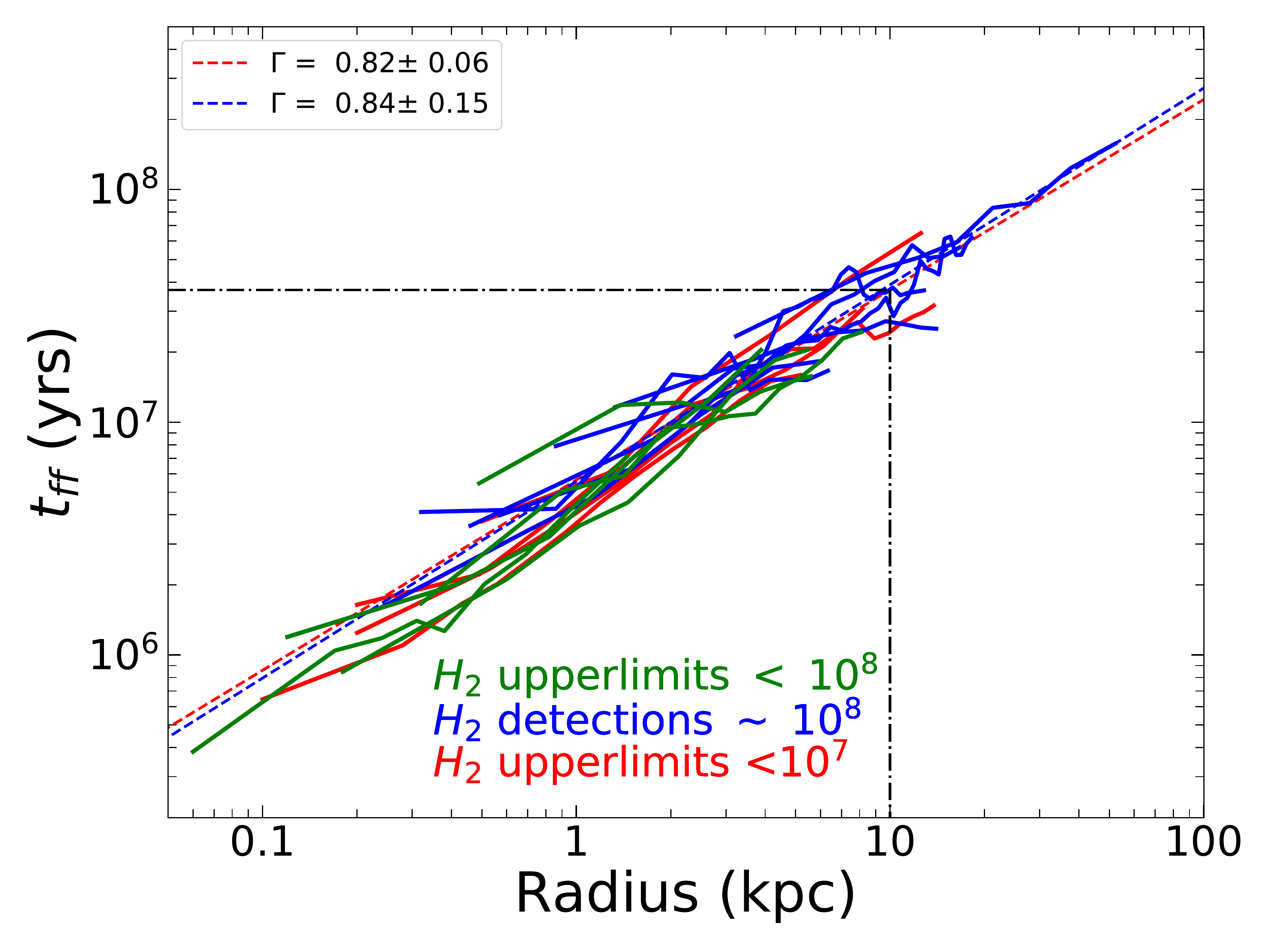}
\caption{The entropy, cooling and free-fall time profiles of low-mass systems with/without evidence of CO detection and H$_2$ gas. The systems, shown in blue, are detected with CO masses exceeding $10^8~\rm M_\odot$ while, shown in red and green, are upper limit detections with CO masses below 10$^7$ and 10$^8$, respectively. Red and blue power law fits (dashed lines) are performed using the BCES routine. Dash-dotted lines correspond to the crossing between the best-fitting lines with observations at 10 kpc.}
\label{fig_mol_w}
\end{figure*}

The presence or absence of a significant molecular gas signal in cluster central galaxies is closely tied to cooling time.  Systems with atmospheric cooling times at roughly 10 kpc in altitude that lie below $\sim 10^9$ yr are likely to be detected \citep{Pulido:17}.  Those with longer cooling times are not.  A similar trend is found using the entropy parameter.   This condition is tested in early-type galaxies in Fig.~\ref{fig_mol_w}. This figure shows that systems with shallower entropy profiles and lower entropy at 10 kpc, shown in blue, are detected with CO masses exceeding $10^8~\rm M_\odot$. Their entropies at 10 kpc are in the range of 10-30 keV cm$^2$, consistent with central galaxies in clusters that are rich in molecular gas. Objects with restrictive CO upper limits that lie at or below $10^7~\rm M_\odot$, shown in red, have entropies lying above 40 keV cm$^2$.  Systems with less restrictive upper limits that lie below $10^8~\rm M_\odot$ have similarly high entropy are shown in green. The entropy profiles presented here are  consistent with \citet{Werner:14}. 

The middle panel of Fig.~\ref{fig_mol_w} shows a similar trend in atmospheric cooling time.   Systems with atmospheric cooling time at 10 kpc that lie below $\sim$ 1 Gyr are detected in CO. This segregation continues inward to approximately 1 kpc, where all profiles merge.  This merging could be real or it may be the consequence of our inability to resolve the gas temperature in the inner bins (see \citet{Hogan:17} for detailed discussion). In fact, there is nothing special about 10 kpc. It is approximately the radius at which the cooling time and entropy parameters can be resolved in cluster centrals. Presumably in those systems, like these, the profiles remain segregated as they move inward. Nevertheless, the profiles show that the atmospheric cooling time in systems with significant molecular gas lie below those without molecular gas over most of the volume within 10 kpc. Like cluster centrals with large molecular gas masses, short cooling times in the atmospheres of normal ellipticals apparently correlate with significant levels of molecular gas. 

Free-fall time profiles are shown in the right-hand panel.  All are similar and reveal little scatter at 10 kpc.  Nevertheless, dividing the cooling time and free-fall time profiles only scatters the trend in Fig.~\ref{fig_tctfprof}.  The free-fall time adds no leverage;   the \tctff\ ratio plays no obvious role. While the sample is admittedly small, this diagram shows that $t_{\rm c}$ is an effective indicator for the presence of cold molecular gas.  Similar trends are found in cluster central galaxies \citep{Hogan:17a, Pulido:17, Babyk17nature}.  This result implicates the hot atmosphere in the production of molecular clouds in early-type galaxies, presumably through cooling.

Finally, three systems in Fig.~\ref{fig_mm} have elevated levels of molecular gas relative to their atmospheric masses.  All three are early spirals and their gas lies in rotating disks.  Molecular disks are often attributed to high-angular momentum mergers, and this may be true.  However, and internal atmospheric origin cannot be ruled out. 

\citet{Werner:14} and \citet{Negri:14} pointed out that thermally unstable cooling may be enhanced in rotating atmospheres. Rotation reduces the effective gravity, and thus prevents a thermally unstable parcel of gas from sinking to its equilibrium position. This would lead to unstable cooling on non-radial orbits. If the cooling gas is unable to shed its angular momentum, it would settle into a molecular disk (see also \citet{Sobacchi:19a}.  Molecular clouds are observed preferentially in fast-rotating early-type galaxies \citep{Davis:19}.  If the atmospheres and stars are co-rotating, thermally unstable cooling may be enhanced, leading to the observed correlation.

\subsection{Merger Origin}

Molecular and atomic gas clouds in early-type galaxies are commonly attributed to mergers.   \citet{Davis:11} compared the angular momentum vectors of the cold gas and stars in ATLAS$^{3D}$ galaxies and attributed their misalignments to oblique-angle mergers.  These studies and others \citep{Young:02, Young:08, Crocker:11} have argued that cold clouds forming from debris lost from stars should share the stellar angular momentum vector.  Misaligned vectors would indicate an external origin. 

Using a simple model prescription incorporating a variety of physical processes including mergers and feedback,  \citet{Davis:19} estimated the gas-rich merger rate of ellipticals in the local Universe. They concluded that mergers are likely the primary source of cold gas in early type galaxies.  In their model, atmospheric cooling, stellar mass-loss, and other sources contribute less to the cold gas budget.  Their conclusion is based in part on the absence of correlation between stellar mass and molecular clouds, as would be expected if cold clouds condensed with from stellar ejecta.   

Misaligned angular momenta need not in themselves imply an external origin for cold gas.  Semi-analytic models of galaxies evolving in a Lambda-CDM cosmology \citep{Lagos:15} indicate that the high fraction of misaligned gas disks in the ATLAS$^{3D}$ catalog may be caused by angular momentum misalignments between cooling hot atmospheres, stars, and dark matter halos. Their simulations further indicated that the frequency of mergers at the low redshifts found in the ATLAS$^{3D}$ sample  $\sim 40\%$ is too low to account for the high incidence of cold gas whose angular momentum is misaligned with the stars. In a related study, \citet{Lagos:14} concluded that most neutral gas (atomic + molecular) in nearby early-type galaxies likely cooled from their hot atmospheres. 
\subsection{Stellar Ejecta}

$Spitzer$ and $Herschel$ observations of early-type galaxies reveal the presence of polycyclic aromatic hydrocarbon, warm and cold H2, and dust (see \citet{Werner:14} and their references). The presence of polycyclic aromatic hydrocarbon and dust in the cold gas clouds indicates that some fraction of the cold gas originated from stellar mass loss \citep{Werner:14}.  However, \citet{Goulding:16} showed that while the thermalisation of stellar ejecta (mass loss, supernovae) is a significant source of atmospheric gas (see also \citet{Pellegrini:18}, their data are inconsistent with a stellar mass-loss origin alone. 


In summary, the origins of cold gas in early-type galaxies is poorly understood. They may acquire their molecular clouds at some level by all three mechanisms.  Occasional mergers would be an appealing mechanism to explain the scatter in the trends presented here.  However, the trends themselves indicate that condensation from hot atmospheres is a significant source of cold gas in early-type galaxies.

\subsection{Neutral Hydrogen in Hot Atmospheres}

This study is focused on molecular gas due to the availability of CO measurements for large numbers of cluster centrals and nearby early-type galaxies. Nevertheless, HI is crucial to the picture.   Studies have shown that early-type galaxies tend to be underabundant in HI compared to spirals  \citep{Young:89, Young:91, Sage:93, Obreschkow:08, Oosterloo:10}.  Although the variance is large, the relatively low average abundance of HI implies a rapid transition of cold atomic clouds to molecular clouds in hot atmospheres. This transition may be related to high atmospheric gas pressures found in giant ellipticals and cluster centrals. 

\citet{Elmegreen:93} showed that HI in diffuse interstellar clouds should transform rapidly into H2 when clouds are shielded from strong ultraviolet radiation and experience high ambient pressure.  Assuming diffuse, non-gravitating clouds have an internal density that scales with ambient pressure, the clouds will quickly transform their HI into H2. \citet{Blitz:04, Blitz:06} examined this conjecture in a sample of nearby disk galaxies.
They found that the ratio of atomic to molecular gas rises from well below unity to nearly 100 as the external pressure rises from $\sim 10^4~ \rm cm^3~K$ to $\sim 10^6~ \rm cm^3~K$. This trend correlates almost linearly with ambient pressure. 

Atmospheres are volume-filling so they are likely to be in pressure contact with molecular clouds.  The atmospheric pressures observed in this sample would place them on the high end of the Blitz-Rosolowsky correlation, consistent with the relatively low abundance of HI in these systems. This issue will be explored in a future work.


\section{Conclusions}\label{sec_conc}

The thermodynamic properties of the hot atmospheres and molecular gas content of 40 early-type galaxies observed with the $Chandra$ observatory and the ATLAS${}^{3D}$ project were analyzed.  Their properties were compared to those of central cluster galaxies rich in molecular gas. Our conclusions are summarized as follows: 

\begin{enumerate}
\item[(i)]  Molecular gas mass is correlated with atmospheric gas mass, atmospheric density, and atmospheric X-ray luminosity over five decades in molecular gas mass, from early-type galaxies to central cluster galaxies. The most distant outliers are early-type disk galaxies, which have higher levels of molecular gas compared to those without molecular disks. 

\item[(ii)] The ratio of cold molecular gas to hot atmospheric gas within 10 kpc of the galaxy is similar over a broad range in halo mass, from giant ellipticals to central cluster galaxies and lies between $10\%$ and $20\%$. 

\item[(iii)] Molecular gas in early-type galaxies if found preferentially in systems when the cooling time of their hot atmosphere lies below $\sim 10^9$ yr at an altitude of 10 kpc. This ``cooling time threshold'' is similar to what is found in cluster central galaxies.  
 


\end{enumerate}

This study indicates a relationship between the molecular gas content of early type galaxies and their hot atmospheres.  The apparent continuity between the molecular gas content and atmospheric properties of cluster central galaxies, which are almost certainly due to cooling,  likewise suggests that some or most of the molecular gas in early type galaxies cooled from their atmospheres. Taking different approaches, \citet{Werner:14} and \citet{Werner:18new} reached similar conclusions.  

Nevertheless, the connection between hot atmospheres and molecular gas in early-type galaxies is not a strong as central galaxies in clusters.  This is largely due to lower gas masses and poorer detection statistics. Additional data from ALMA and other cold gas tracers are needed to further explore this conjecture.  



\acknowledgements
 
BRM  acknowledge support from the Natural Sciences and Engineering Research Council of Canada. HRR acknowledges support from an STFC Ernest Rutherford Fellowship. ACE acknowledges support from STFC grant ST/P00541/1. BRM thanks Leo Blitz for an insightful discussion during a visit to the Flatiron Institute.  We thank the referee for comments that significantly improved the paper. The scientific results reported in this article are based on observations made by the $Chandra$ X-ray Observatory and has made use of software provided by the $Chandra$ X-ray Center (CXC) in the application packages CIAO, ChIPS, and Sherpa. This paper makes use of the following ALMA data: \#2013.1.00229.S, \#2015.1.00989.S, \#2015.1.01198.S, and \#2015.1.01572.S. ALMA is a partnership of ESO (representing its member states), NSF (USA) and NINS (Japan), together with NRC (Canada), MOST and ASIAA (Taiwan), and KASI (Republic of Korea), in cooperation with the Republic of Chile. The Joint ALMA Observatory is operated by ESO, AUI/NRAO and NAOJ.

\bibliographystyle{apj}
\bibliography{paper}

\end{document}